\documentclass[fleqn,usenatbib]{mnras}

\usepackage{newtxtext,newtxmath}

\usepackage[T1]{fontenc}

\DeclareRobustCommand{\VAN}[3]{#2}
\let\VANthebibliography\thebibliography
\def\thebibliography{\DeclareRobustCommand{\VAN}[3]{##3}\VANthebibliography}

\usepackage{graphicx}	
\usepackage{amsmath}	
\usepackage{xcolor}     
\usepackage{subfig}
\usepackage{supertabular}
\usepackage{longtable}
\usepackage{tabularx}
\usepackage{graphicx}
\usepackage{lscape}
\usepackage{adjustbox}
\usepackage{caption}

\usepackage{listings}
\usepackage{color}
\definecolor{dkgreen}{rgb}{0,0.6,0}
\definecolor{gray}{rgb}{0.5,0.5,0.5}
\definecolor{mauve}{rgb}{0.58,0,0.82}

\title[ALMA counterparts of SDSS quasars]{ALMA High-Level Data Products: Submillimetre counterparts of SDSS quasars in the ALMA footprint}

\author[A. Wong et al.]{
A. Wong$^{1,2}$,
E. Hatziminaoglou$^{1}$\thanks{E-mail: ehatzimi@eso.org},
A. Borkar$^{3}$,
G. Popping$^{1}$,
I. P\'erez-Fournon$^{4,5}$,
F. Poidevin$^{4,5}$,
\newauthor
F. Stoehr$^{1}$,
H. Messias$^{6,7}$
\\
$^{1}$ESO, Karl-Schwarzschild-Str. 2, 85748 Garching bei M\"unchen, Germany\\
$^{2}$Nanyang Technological University, 50 Nanyang Avenue, 639798 Singapore \\
$^{3}$Astronomical Institute, Czech Academy of Sciences, Bocn\'i II 1401, 141 00 Prague, Czech Republic\\
$^{4}$Instituto de Astrof\'{i}sica de Canarias, E-38205 La Laguna, Tenerife, Spain \\
$^{5}$Departamento de Astrof\'{i}sica, Universidad de La Laguna, E-38206 La Laguna, Tenerife, Spain \\
$^{6}$Joint ALMA Observatory, Alonso de C\'ordova 3107, Vitacura 763-0355, Santiago, Chile \\
$^{7} $European Southern Observatory, Alonso de C\'ordova 3107, Vitacura, Casilla 19001, Santiago de Chile, Chile
}

\date{Accepted XXX. Received YYY; in original form ZZZ}

\pubyear{2022}

\begin{document}
\label{firstpage}
\pagerange{\pageref{firstpage}--\pageref{lastpage}}
\maketitle

\begin{abstract}
The Atacama Large Millimetre/submillimetre Array (ALMA) is the world's most advanced radio interferometric facility, producing science data with an average rate of about 1 TB per day. After a process of calibration, imaging and quality assurance, the scientific data are stored in the ALMA Science Archive (ASA), along with the corresponding raw data, making the ASA an invaluable resource for original astronomical research. Due to their complexity, each ALMA data set has the potential for scientific results that go well beyond the ideas behind the original proposal that led to each observation. For this reason, the European ALMA Regional Centre initiated the High-Level Data Products initiative to develop science-oriented data products derived from data sets publicly available in the ASA, that go beyond the formal ALMA deliverables. The first instance of this initiative is the creation of a catalogue of submillimetre (submm) detections of Sloan Digital Sky Survey (SDSS) quasars from the SDSS Data Release 14 that lie in the aggregate ALMA footprint observed since ALMA Cycle 0. The ALMA fluxes are extracted in an automatic fashion, using the ALMA Data Mining Toolkit. All extractions above a signal-to-noise cut of 3.5 are considered, they have been visually inspected and the reliable detections are presented in a catalogue of 376 entries, corresponding to 275 unique quasars. Interesting targets found in the process, i.e. lensed or jetted quasars as well as quasars with nearby submm counterparts are highlighted, to facilitate further studies or potential follow up observations.
\end{abstract}




\section{Introduction}
\label{sec:intro}

Scientific archives are not only the legacy of their facilities, guaranteeing the fundamental scientific principle of the reproducibility of the results, but a primary resource for new, independent research. Archival data complement present-day research, as they provide the foundation on which knowledge can be progressively built on. As the interpretation of information evolves over time, thanks to improved or new methodologies, archive contents allow for the re-evaluation of earlier data sets and results, with the capacity to influence the research done in the present. All the above are especially true for data sets that are as rich as those produced by the Atacama Large Millimetre/submillimetre Array (ALMA), where the original publication by the Principal Investigator (PI) often enough only partially exploits the full potential of the entire data set. While valid for each individual data set, this statement is particularly true for the combined body of all data sets that have been observed by a facility over time. That combined body forms a new data set in its own right enabling cross-project research, such as statistical analyses or deep data combination. 

\begin{figure}
\centering
\includegraphics[angle=-90,width=8.5cm]{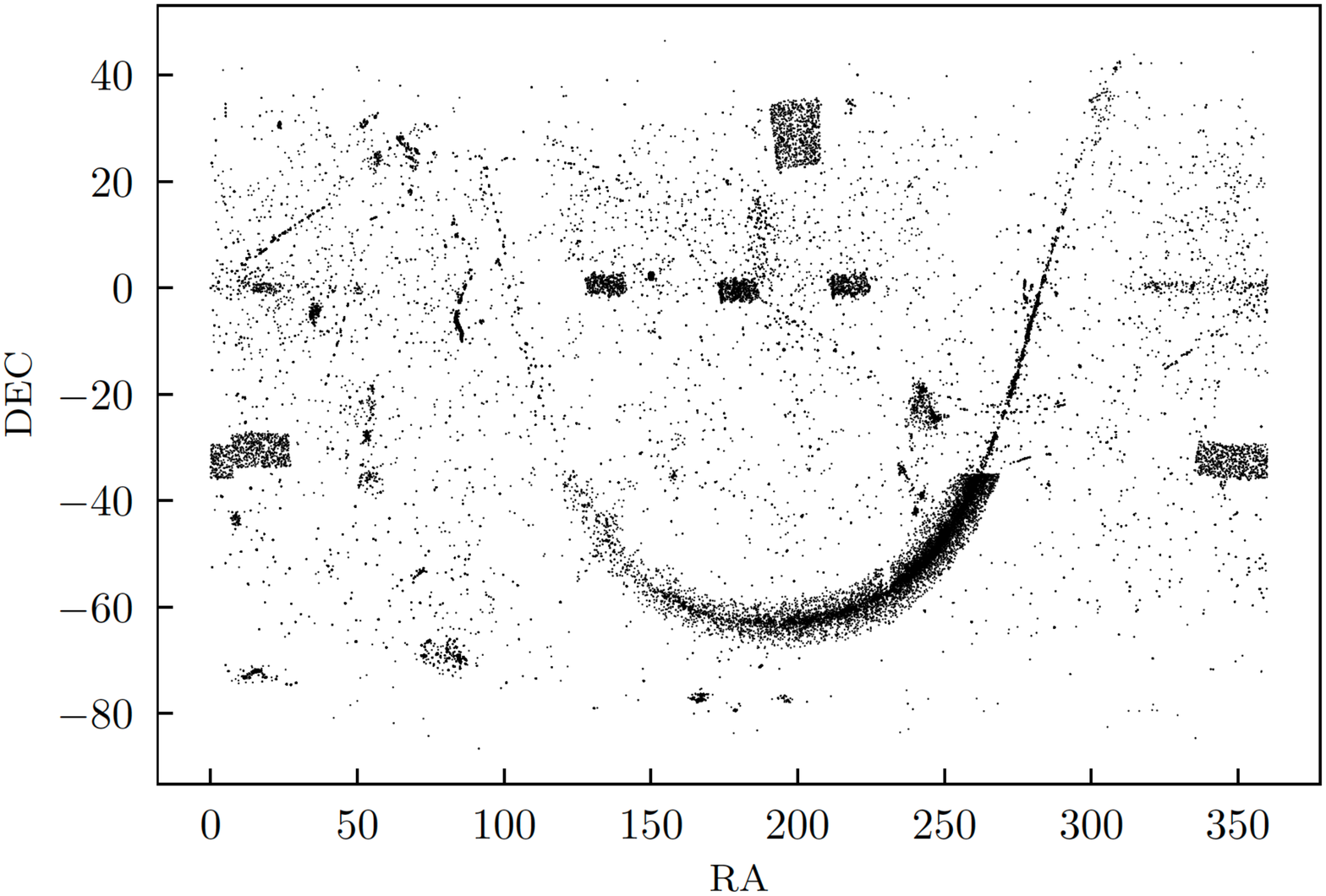}
\caption{The ALMA science target observations footprint (calibration fields are not included). Note that, in most cases, the ALMA field of view is smaller than the pixel.}
\label{fig:almafootprint}
\end{figure}

\begin{figure}
\centering
\includegraphics[width=8.5cm]{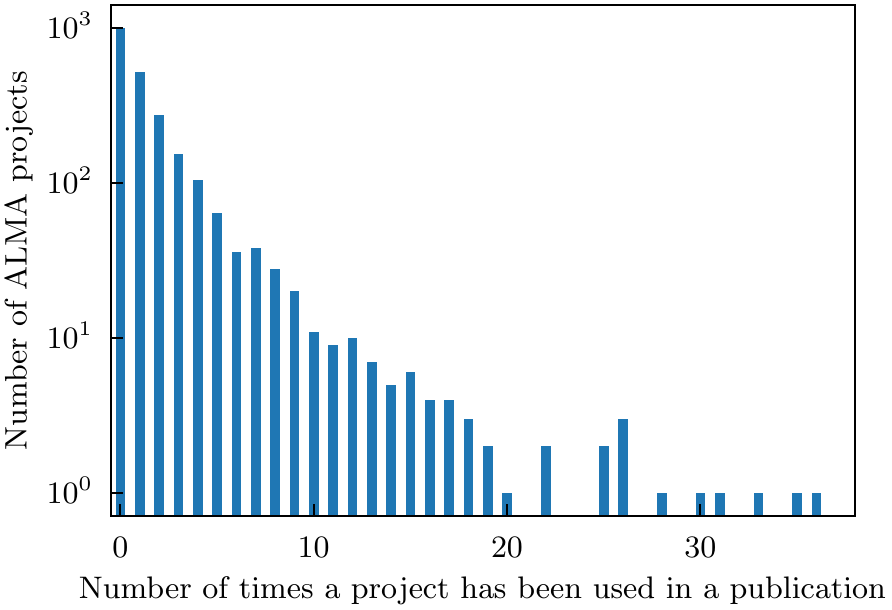}
\caption{Re-usage statistics of ALMA projects.}
\label{fig:reusage}
\end{figure}

After a decade of observations, the ALMA Science Archive (ASA) contains more than 19000 Member Observation Unit Sets (MOUSs). An MOUS roughly corresponds to a Science Goal defined in a proposal at the time of submission, split per array, configuration or spectral tuning. Here we define the ``ALMA footprint'' as the aggregate region of the sky with at least one ALMA exposure (i.e. MOUS) that has been observed as part of PI observations since the start of ALMA operations (Cycle 0) all the way to Cycle 8, including all observations with proprietary time that expired by November 1, 2022. The ALMA footprint is shown in Fig. \ref{fig:almafootprint}. ALMA is currently producing between 300 and 400 TB of raw and reduced data per year. The potential for archival research is steadily increasing with time as more observations get added. This is also evidenced by the raising fraction of ALMA publications that make use of ALMA archival data: In 2021, 28\% of all ALMA peer-reviewed publications were (partially) based on archival data \citep{stoehr22}. ALMA data are used over and over again to produce original science, as shown in Fig. \ref{fig:reusage}, with some ALMA projects having been used in more than 20 publications. At the same time, the over-subscription for ALMA PI observations has never been higher, with the pressure reaching above 7 in Cycle 9 in Europe\footnote{https://almascience.eso.org/documents-and-tools/cycle9/cycle-9-proposal-submission-statistics}. Therefore, while it is increasingly difficult to obtain PI data due to the over-subscription, the potential of conducting archival research is growing. The usage of ALMA archival data is supported by the ALMA Regional Centres (ARCs) in Europe, North America and East Asia, that continuously assist ALMA users also through face-to-face visits, in making optimal use of the wealth of the ASA to conduct original, high-impact research.

One of the key strategic goals of ALMA is to further increase the scientific usage of the ASA\footnote{https://www.eso.org/sci/facilities/alma/announcements/20180712-alma-development-roadmap.pdf} and the development of science-oriented, curated data products to further facilitate the scientific exploitation of the ASA. In response to this, the European ARC network \citep{hatzimi15} launched the High-Level Data Products (HLDP) initiative. The goal of the initiative is to develop such science-oriented data products of interest to the community and to develop, internally, the knowledge to support the community with a diverse range of archival science cases. The current efforts within the HLDP initiative include the creation of new image products and catalogues based on information stored in the ASA, as well as tools that facilitate the interaction with the ASA.

As a demonstration of the  capabilities of HLDP-like initiatives, we present in this paper a catalogue of ALMA detections of Sloan Digital Sky Survey (SDSS) quasars within the ALMA footprint. 
The paper is structured as follows: Section \ref{sec:sdsssample} describes the SDSS quasar sample in the ALMA footprint. Section \ref{sec:extraction} delineates the methodology of the extraction of the ALMA fluxes from the fits products in the ASA and subsequent visual inspection. The astroquery script used for the identification of SDSS quasars in the ALMA footprint is given in Appendix \ref{sec:astroquery}. Section \ref{sec:submmemission} discuses the contents of the catalogue of submillimetre (submm) counterparts of the SDSS quasars. Section \ref{sec:notes} presents sources of particular interest found during the analysis, i.e. jetted and lensed quasars as well as objects with nearby submm companions, while Sec. \ref{sec:discuss} summarises the outcomes of this work and indicates how quasar research could benefit from such data sets. All the tables produced by this work are presented in Appendix \ref{sec:catalogues}. 

\section{SDSS quasars in the ALMA footprint}
\label{sec:sdsssample}

The parent sample for the present study is the SDSS Data Release 14 (DR14) catalogue of AGN and quasars \citep[DR14Q;][]{paris18}. The DR14Q consists of 526,356 spectroscopically confirmed quasars from all SDSS data releases up until DR14, which have confirmed redshift, spectral properties and black hole masses \citep{rakshit20}. Setting aside sources with a declination above 40 deg, that are not observable by ALMA, results in a list of 357,392 objects to be considered.

These sources were matched against the metadata of the ASA to identify the objects within the ALMA footprint, among observations taken under scientific categories ``Active galaxies'', ``Cosmology'', ``Galaxy evolution'' and ``Local Universe'', using the script presented in Appendix \ref{sec:astroquery}. Categories ``Disks and planet formation'', ``ISM and Star Formation'', ``Stars and Stellar Evolution'' and ``Solar System'' were left out of this pilot project, as many of these have complex continuum images with overlapping structures. The aim of the work presented here was to put together an automated methodology that has the potential to be repeated in regular intervals, with minimum requirements for visual inspection of every single extraction. Future attempts for an updated catalogue may include images from the remaining science categories, too. The SDSS quasars may have been the primary target of each observation or they may be lying in a region observed by ALMA while targeting different objects. The match resulted in a total of 1295 entries, i.e. 1295 instances of SDSS quasars falling within the ALMA footprint. This corresponds to 732 unique sources within the regions covered by 288 unique ALMA projects, with a total of 790 MOUSs.

Note that, in addition to the objects presented here, hundreds of DR14Q quasars have been observed by ALMA as calibrators. The analysis of the submm continuum emission of these calibrators has been discussed in \citealt{bonato18} and are not included in this work.

\section{Extraction of submm counterparts}
\label{sec:extraction}

To extract the submm continuum counterparts of SDSS quasars in the ALMA footprint in an automated fashion, we made use of the ALMA Data Mining Toolkit \citep[ADMIT\footnote{https://admit.astro.umd.edu/};][]{teuben15}. ADMIT is a python-based execution environment and set of tools designed for analysing image data cubes. Running ADMIT was not limited to the region around the SDSS coordinates but it was run, instead, on the full images. To search for continuum emission we made use of the task \texttt{Sfind2D\_AT}, which is based on the CASA \texttt{findsources} task. This task searches for point-like sources in continuum images, by identifying emission peaks brighter than $N$ times the typical 1 $\sigma$ noise (signa-to-noise or SNR) limit of the data, where $N$ is pre-defined by the user. In this work we require that the emission peaks have a flux of at least $N$ = 3.5 times the 1 $\sigma$ noise limit of the data. Once an emission peak is identified, the \texttt{Sfind2D\_AT} task then fits the region around the brightest pixel value with a two-dimensional Gaussian to obtain integrated flux and shape (major and minor axis size and position angle) information. In general the \texttt{Sfind2D\_AT} task does not recover highly resolved sources with complex structures well (e.g., rings or arms), breaking these down into multiple (sometimes overlapping) Gaussians. These cases were assessed at the time of the visual inspection. 

The initial search for a bright peak is performed on primary-beam uncorrected images to ensure a coherent noise profile throughout the image. The Gaussian fit and flux determination is performed on the primary-beam corrected image. In the few cases where no primary-beam information was available for an MOUS in the ASA (this can happen for some data taken during the early ALMA observing cycles), the source finding and fitting procedures were both performed on the primary-beam corrected image available in the ASA. The final ADMIT output consists of a full list of sources detected within each ALMA image that contains coordinates of at least one SDSS quasar. 

To identify the submm source associated with each of the SDSS quasars, we cross-matched the ADMIT list with the reduced DR14Q catalogue, using a matching radius of 5$\arcsec$ centred on the SDSS coordinates. All of the extraction within this 5$\arcsec$ radius were then visually inspected. 

\subsection{Visual inspection}
\label{sec:visual}
The veracity of the ADMIT extractions was determined through visual inspection. A total of 1707 ALMA extractions were returned after the cross-matching between the ADMIT list and DR14Q catalogue, applying a 5$\arcsec$ search radius around the SDSS coordinates. All ALMA images 
containing one of these extractions were inspected using CARTA\footnote{https://cartavis.org}, to identify image artefacts, unreliable extractions or extractions that did not correspond to the SDSS quasar. 

\begin{figure*}
\captionsetup[subfloat]{labelformat=empty}
  \subfloat[{(a)}]{
	\begin{minipage}[c][1\width]{
	   0.3\textwidth}
	   \centering
	   \includegraphics[width=4.5cm]{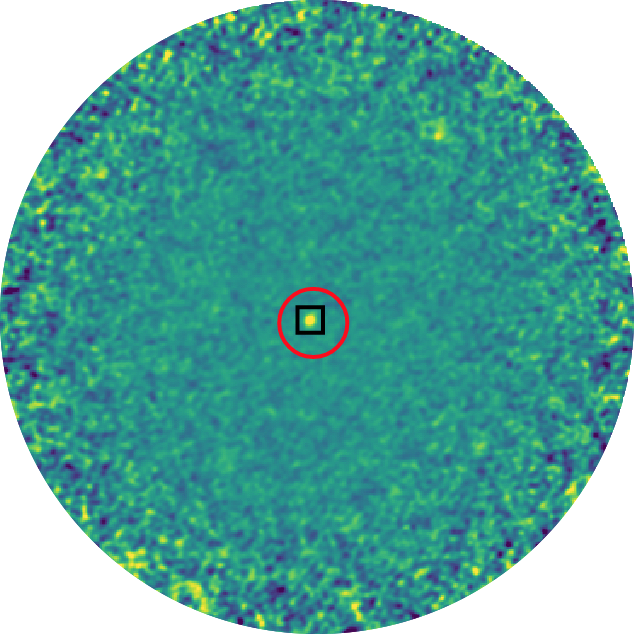}
	\end{minipage}}
  \subfloat[{(b)}]{
	\begin{minipage}[c][1\width]{
	   0.3\textwidth}
	   \centering
	   \includegraphics[width=4.5cm]{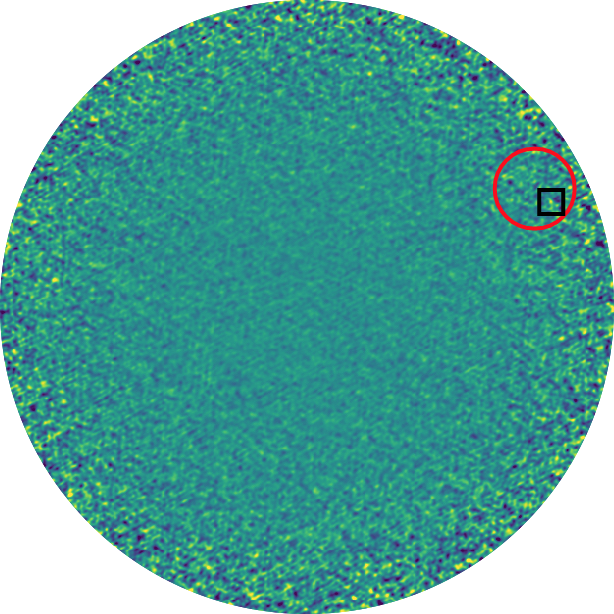}
	\end{minipage}}
  \subfloat[{(c)}]{
	\begin{minipage}[c][1\width]{
	   0.3\textwidth}
	   \centering
	   \includegraphics[width=4.5cm]{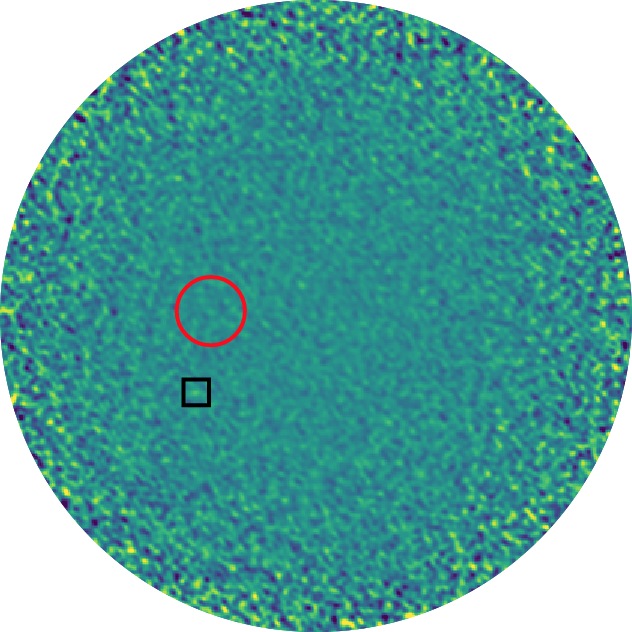}
	\end{minipage}}
	
    \subfloat[{(d)}]{
	\begin{minipage}[c][1\width]{
	   0.3\textwidth}
	   \centering
	   \includegraphics[width=4.5cm]{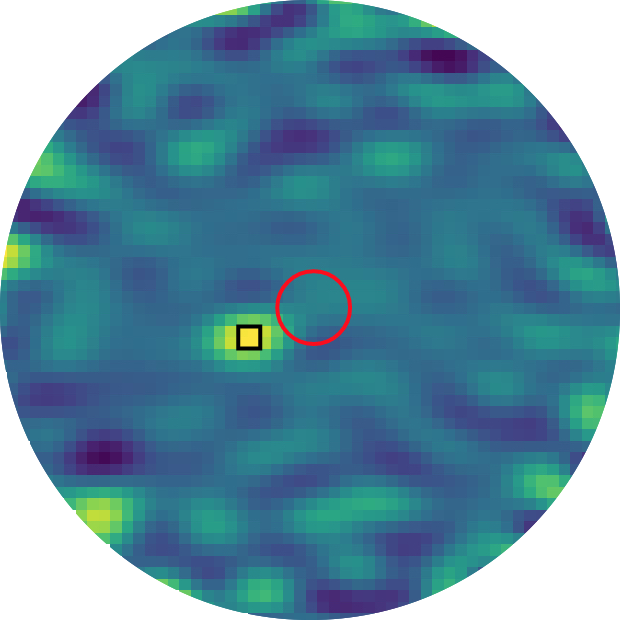}
	\end{minipage}}	
  \subfloat[{(e)}]{
	\begin{minipage}[c][1\width]{
	   0.3\textwidth}
	   \centering
	   \includegraphics[width=4.5cm]{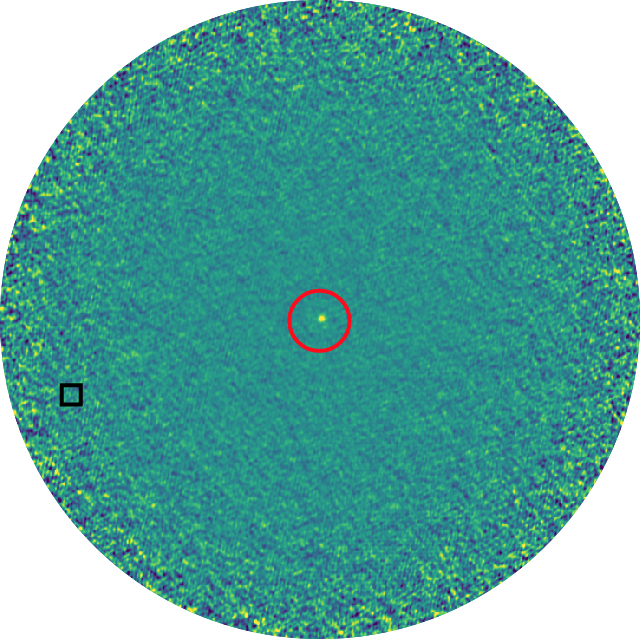}
	\end{minipage}}
\caption{Examples of cases encountered during the visual inspection of ALMA images, performed to determine the veracity of the ADMIT extractions. The black square and red circle correspond to the coordinates of the submm ALMA source and the SDSS quasar respectively. The circles are (a) SDSS and ALMA counterparts coinciding within 1.5$\times \alpha^b_{max}$. (b) and (c) illustrate cases of spurious ALMA extractions. (d) Shows the case of a submm source extraction in the vicinity of the SDSS quasar while the quasar itself does not emit in the submm. Finally, (e) shows a missed ADMIT extraction, with a visible submm source at the exact location of the SDSS quasar. The sizes of the circles and squares are 4\arcsec and 2\arcsec across respectively.}
\label{fig:VisualInspectionCases}
\end{figure*}

Figure \ref{fig:VisualInspectionCases} showcases the main scenarios that were encountered during visual inspection. Example (a) shows a confirmed submm counterpart to an SDSS quasar. Of the 1707 extractions, 1225 were deemed non-reliable ALMA counterparts of the SDSS quasars because of one of the following three cases: $i)$ 107 extractions, corresponding to 25 unique objects, occurring in the outer and noisier parts of the ALMA images (see e.g. Fig. \ref{fig:VisualInspectionCases} panels (b) and (c), showing non-reliable extractions coinciding (or not) with the location of the quasar); $ii)$ while no submm counterpart was visible at the location of the quasar, an ADMIT extraction occurred in the vicinity of the quasar (for example Fig. \ref{fig:VisualInspectionCases}, panel (d));  and $iii)$ four occasions (four unique objects) for which, while an obvious ALMA counterpart was located right at the position of the SDSS quasar, no automatic flux extraction was performed (see Fig. \ref{fig:VisualInspectionCases}, panel (e) for an example). This was due to the lack of a primary beam response curve associated to the corresponding images discussed in the previous section. In these cases, the (high) noise level at the outer edges of the image dominated the noise level of the whole image, causing sources that were clearly visible on the images to fail the SNR cut of 3.5. These four sources were not included in Table \ref{tab:main} but we provide their information separately, in Table \ref{tab:pbc}. 

\begin{figure}
\centering
\includegraphics[width=8cm]{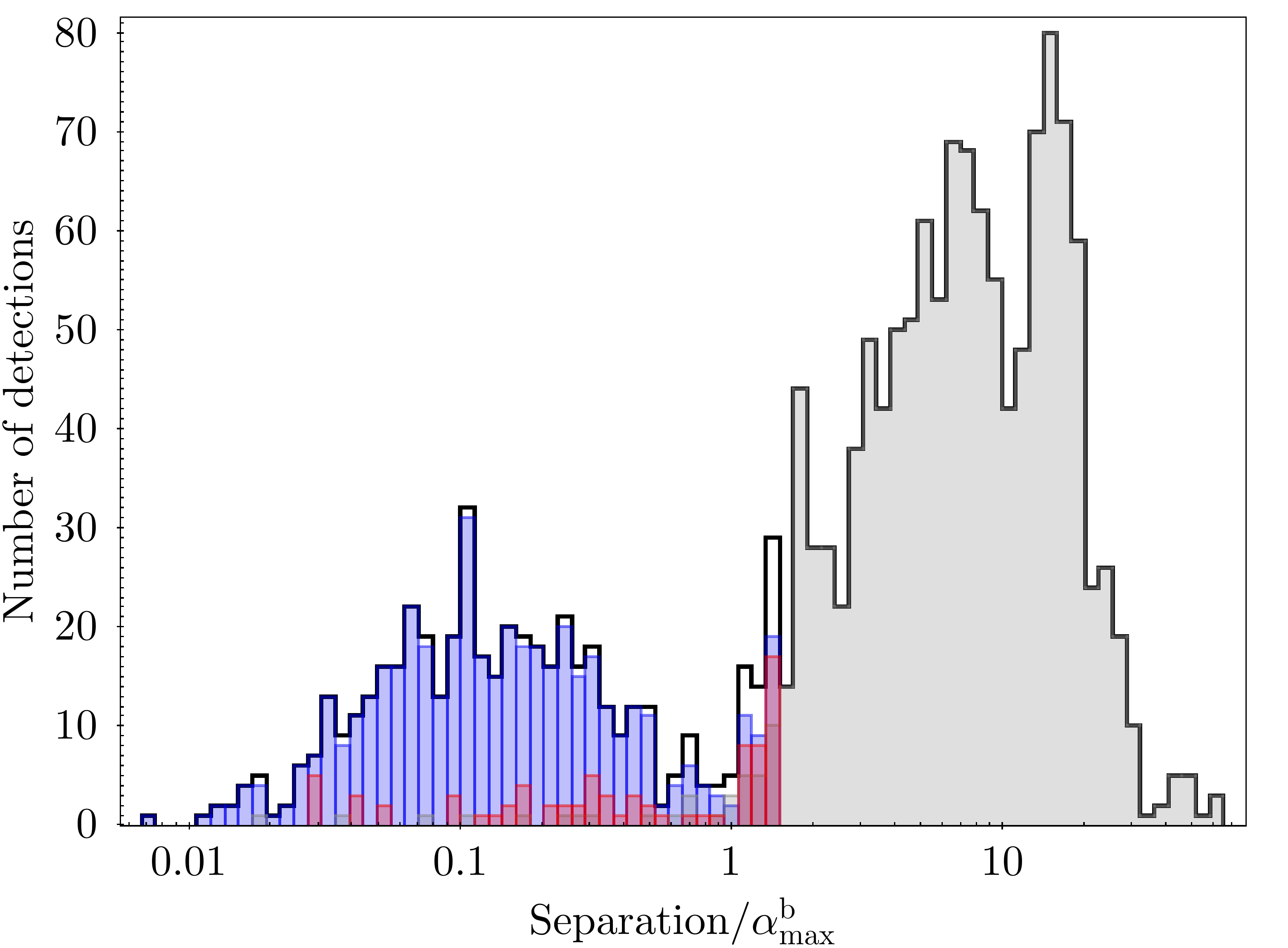}
\caption{Distribution of the ratio between the separation of the SDSS coordinates from the ALMA coordinates and $\alpha^b_{max}$ for each extraction, for the reliable extractions (blue), the multiple extractions occurring on the same sources (red) and the non-reliable detections (grey). The histogram in black is the sum of all.}
\label{fig:sepalphahisto}
\end{figure}

The best differentiator found to distinguish between reliable and unreliable detection was the ratio of the separation between the optical and submm coordinates and the beam major axis, $\alpha^b_{max}$, as shown in Fig.  \ref{fig:sepalphahisto}. Visual inspection indicated that a submm extraction was associated to the SDSS quasar if the ALMA and SDSS coordinates coincide within 1.5$\times \alpha_{max}^{b}$, a value slightly above the minimum of $\sim$1 seen in the bimodal distribution of this quantity in Fig. \ref{fig:sepalphahisto}. Only 2.5\% of the non-reliable detections (filled grey histogram) have a ratio below 1.5, with the fraction dropping to 1\% when a cut at a ratio of 1.0 is applied. At the same time, only 1.5\% of all reliable detections (blue histogram) have a value above 1.0. In other words, in order to fully automate the method without further need for visual inspection, a cut at Separation / $\alpha^b_{max}$ = 1.0 could be imposed, ensuring minimum contamination and maximum completeness of the resulting catalogue.

For some cases (122 in total, corresponding to 34 unique quasars), ADMIT performed multiple extractions on the same source (red spikes in Fig. \ref{fig:sepalphahisto}. In these cases, the extraction centred closer to the SDSS coordinates was kept. In the few (six) cases where two extractions were performed on exactly the same point in the image, the extraction with the highest peak SNR was kept. The multiple extractions are shown in red spikes in Fig. \ref{fig:sepalphahisto}. 36\% of those cases have Separation / $\alpha^b_{max} > 1.0$. The remaining ones, for a full automation of the process, can be removed by simply applying the minimum separation and peak SNR criteria described above.

\section{The ALMA counterparts of SDSS quasars}
\label{sec:submmemission}

\begin{figure}
\centering
\includegraphics[width=8cm]{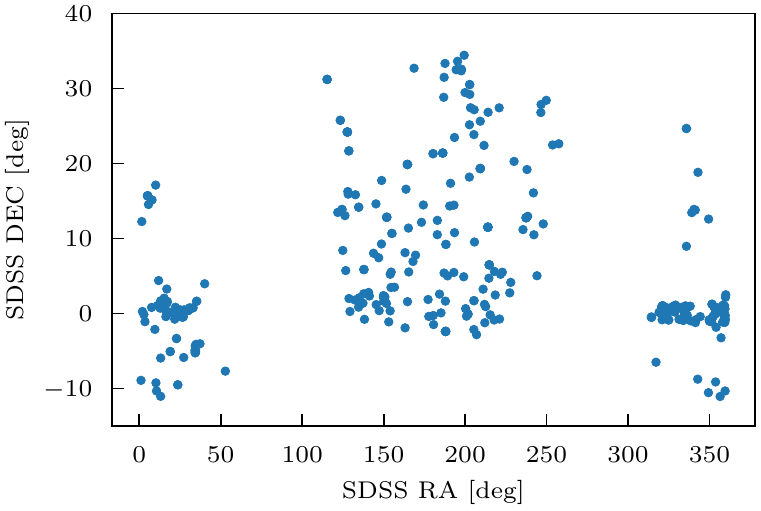}
\caption{SDSS coordinates of quasars with submm counterparts in the ALMA footprint.}
\label{fig:sdssradec}
\end{figure}

\begin{figure}
\centering
\includegraphics[width=8cm]{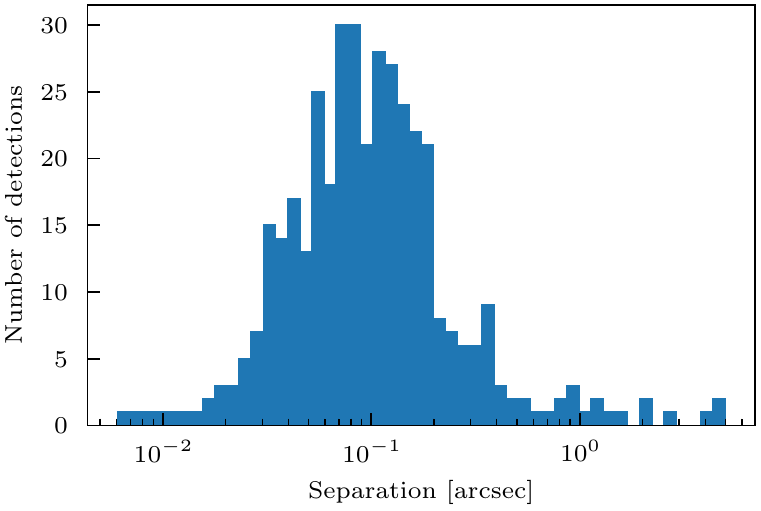}
\caption{The distribution of the separation between the optical (SDSS) and submm (ALMA) coordinates of the quasars with submm counterparts confirmed by visual inspection.}
\label{fig:separations}
\end{figure}

Following the visual inspection, the submm counterparts of SDSS quasars in the DR14Q catalogue were identified. A total of 376 reliable detections of 275 unique objects from 84 unique projects were found, from observations taken between July 18, 2014 and September 25, 2021, and with a proprietary time that expired before November 1, 2022. Of those 376 observations, 87\% were targeting the SDSS quasar (i.e. the observations are centred on the quasar), while for the remaining cases the quasar just happened to be on an ALMA image centred on a different target. Lensed quasars are not included in these numbers, as they break into multiple submm detections  but are discussed separately in Sec. \ref{sec:lenses}.

\begin{figure}
    \centering
    \includegraphics[width=7.5cm]{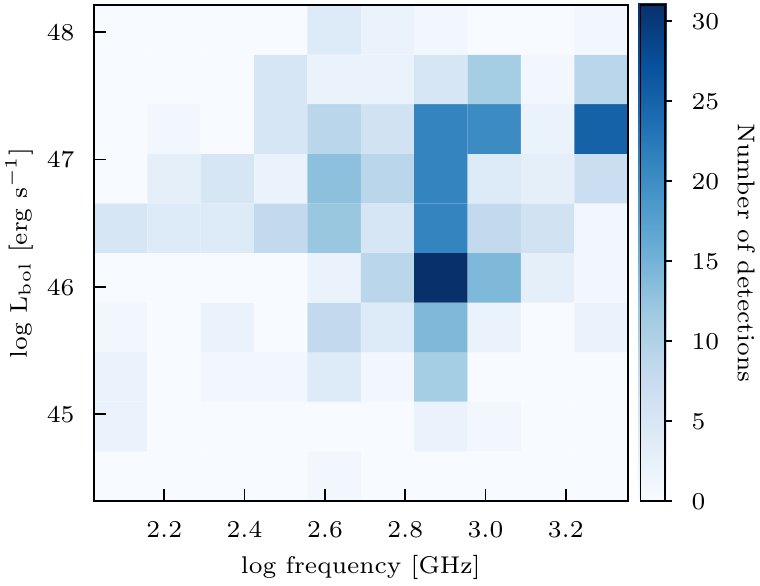}
    \caption{Distribution of the L$_{bol}$ as a function of the rest frequency, corresponding to the representative frequency of each band shifted to $z=0$,  for all the submm counterparts of the SDSS quasars.}
    \label{fig:lbolfreq}
\end{figure}

The sky distribution of the SDSS quasars with ALMA counterparts identified in this work and the distribution separations (in arcsec) between the SDSS and ALMA positions are shown in Figs. \ref{fig:sdssradec} and \ref{fig:separations}, respectively. 
Fig. \ref{fig:lbolfreq} shows the distribution of the bolometric luminosities, L$_{bol}$, of quasars with submm detections as a function of rest frequency of the submm observations, corresponding to the representative frequency of each ALMA band shifted to $z=0$. The figure provides a representation of the parameter space sampled by the continuum observations with ALMA of the SDSS quasars with L$_{bol}$ between 10$^{45}$ and 10$^{48}$ erg/s, indicating that most of the observations targetted the star formation in the hosts of quasars at $\sim$790 GHz rest frequency (or $\sim$320 micron).

\begin{figure}
    \centering
    \includegraphics[width=7.5cm]{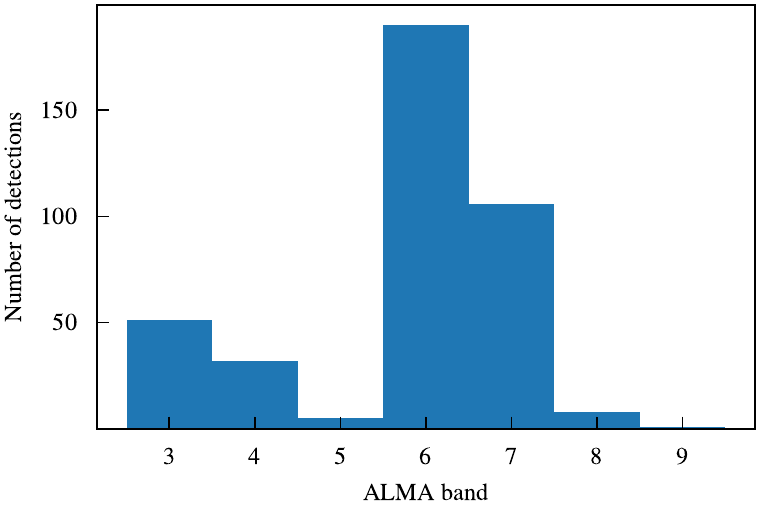}
    \caption{Number of submm counterparts of SDSS quasars per band.} 
    \label{fig:bands}
\end{figure}

ALMA counterparts of SDSS quasars were found in  observations carried out in bands 3 to 9, with the band distribution shown in Fig. \ref{fig:bands}. Most of the observations were done in bands 3, 6 and 7, with 51, 190 and 106 observations, respectively. 6\% (2\%) of the objects had ALMA counterparts in two (three) bands and one object was observed in four ALMA bands. 

\begin{figure}
    \centering
    \includegraphics[width=8cm]{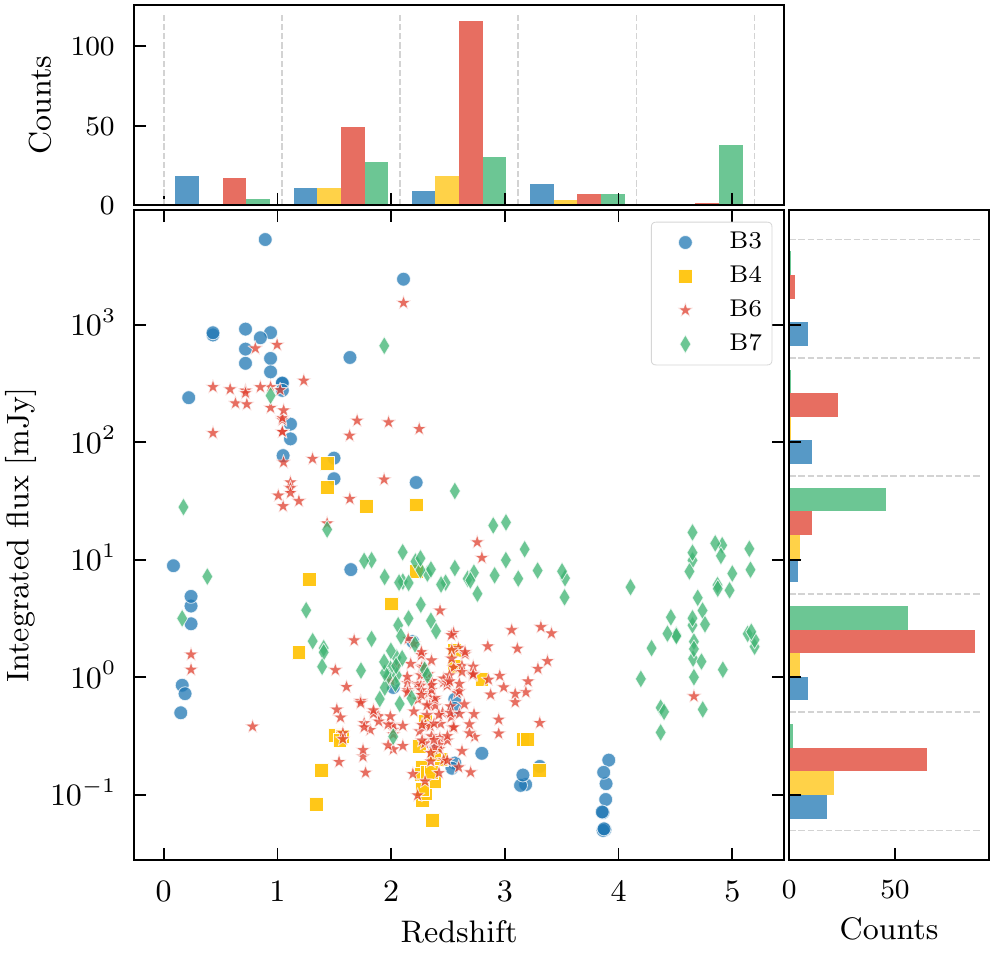}
    \caption{Integrated flux per band as a function of redshift, for the four bands with the larger number of ALMA observations. The histograms show the number of sources in each redshift bin (top histogram) and in each flux bin (right histogram). The dotted lines mark the bin edges.}
    \label{fig:intflux}
\end{figure}

Fig. \ref{fig:intflux} presents the integrated flux as a function of redshift, for bands 3 (blue circles), 4 (yellow squares), 6 (red stars) and 7 (green diamonds). The adjacent histograms show the distribution of the two parameters. 70\% of the objects are in the redshift range between 1 and 3, encompassing the peak of quasar activity. The sample is in no way complete or statistical and any trends seen in this figure are not representative of the quasar population. The bright fluxes in bands 3 and 4 (above a couple tens of mJy) are indicative of synchrotron emission out to redshifts of $\sim$2, while the bulk of bands 6 and 7 emission (at redhifts above $\sim$1.3) traces star formation in the hosts of the quasars.

The ALMA submm counterparts of DR14Q quasars were compiled into a catalogue, an excerpt of which is shown in Table \ref{tab:main}. The table contains the SDSS object name and coordinates; the optical redshift; the ALMA source coordinates; the separation between the two sets of coordinates; the ALMA band and array (12-m or 7-m, the latter including TP observations); the angular resolution and continuum sensitivity of the image from which the submm counterpart was extracted; the peak and integrated fluxes with their respective errors; the major and minor axes of the extracted ellipse and associated  position angle; the beam major and minor axes and position angle; the image RMS; the date of observation; the project code and MOUS ID; and a column indicating whether an entry is among with ALMACAL objects. Indeed, of the 276 unique quasars in the catalogue, 19 are in common with the 754 calibrators in the ALMACAL catalogue \citep{bonato18}. 54 unique objects were observed more than once in the same band as part of different projects. In such cases, all detections are included in the table. The full catalogue is available as online material.

Non-detection information has a scientific value, when e.g. looking for detection limits. For this reason, we compiled the list of DR14Q quasars within the ALMA footprint for which no detection above an SNR cut of 3.5 was found on any of the images that contained them. The objects are listed in Table \ref{tab:NonDetected}. Other than the SDSS ID, coordinates and redshift, the table includes the ALMA band and array, the angular resolution and continuum sensitivity, the project code and the MOUS ID of the data set including the individual quasars' coordinates.

\section{Notes on individual objects}
\label{sec:notes}

The visual inspection of the list of SDSS quasars with ALMA counterparts revealed a number of interesting sources, presented in this section, namely jetted quasars (Sec. \ref{sec:jets}), lensed quasars (Sec. \ref{sec:lenses}) and quasars with more than one submm counterparts (physical or projected) within 5$\arcsec$ (Sec. \ref{sec:companions}). Each object is described briefly and CARTA cutouts for each object are shown in Figs. \ref{fig:jets}, \ref{fig:lens} and \ref{fig:companions}, for jetted, lensed and objects with multiple submm counterparts, respectively.

\subsection{Jetted quasars}
\label{sec:jets}

\begin{figure*}
\captionsetup[subfloat]{labelformat=empty}
  \subfloat[{SDSS\_J002025.22+154054.7 [3]}]{
	\begin{minipage}[c][1\width]{
	   0.3\textwidth}
	   \centering
	   \includegraphics[width=4.5cm]{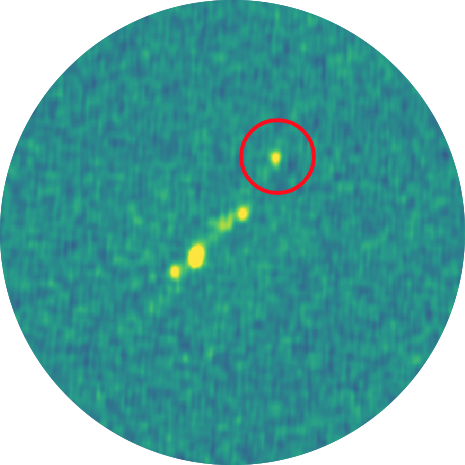}
	\end{minipage}}
  \subfloat[{SDSS\_J085841.44+140944.8 [3]}]{
	\begin{minipage}[c][1\width]{
	   0.3\textwidth}
	   \centering
	   \includegraphics[width=4.5cm]{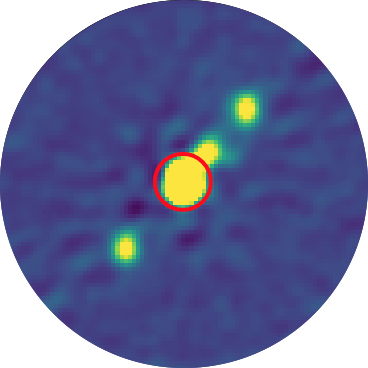}
	\end{minipage}}
  \subfloat[{SDSS\_J123200.01-022404.7 [6]}]{
	\begin{minipage}[c][1\width]{
	   0.3\textwidth}
	   \centering
	   \includegraphics[width=4.5cm]{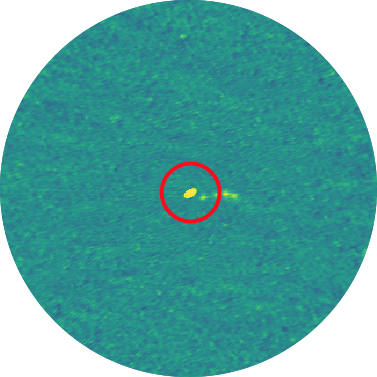}
	\end{minipage}}

  \subfloat[{SDSS\_J135706.53+253724.4 [4]}]{
	\begin{minipage}[c][1\width]{
	   0.3\textwidth}
	   \centering
	   \includegraphics[width=4.5cm]{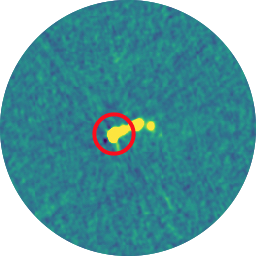}
	\end{minipage}}	
  \subfloat[{SDSS\_J170955.01+223655.7 [4]}]{
	\begin{minipage}[c][1\width]{
	   0.3\textwidth}
	   \centering
	   \includegraphics[width=4.5cm]{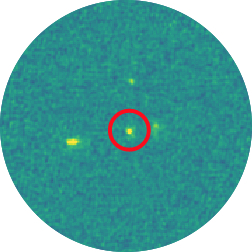}
	\end{minipage}}	
\caption{ALMA cutouts of the jetted SDSS quasars in the ALMA footprint. The location of the optical quasar is indicated with a red circle. The corresponding ALMA band is indicated in square brackets next to the name of each object. All cutouts measure 20$^{\prime\prime}$ in diameter while the red circles are 3\arcsec across.}

\label{fig:jets}
\end{figure*}

Five SDSS quasars revealed submm jets in the ALMA images, shown in Fig. \ref{fig:jets}. For these objects, the main catalogue (Table \ref{tab:main}) only includes the fluxes of the nucleus - not those from the various parts of the jet. Some details about the jetted quasars are given below.

SDSS\_J002025.22+154054.7 (3C 9) is a radio-loud quasar at a redshift of 2.019, with radio and X-ray jets pointing to the north-west and to the south-east of the quasar. The X-ray emission of this object exceeds the radio emission by a factor five. Several mechanisms have been suggested to explain the X-ray emission, namely non-thermal synchrotron, non-thermal inverse Compton and thermal emission \citep{fabian03}. Compared to the 4.9\,GHz map \citep{bridle94}, only the knots in the south-east jet are detected on the ALMA band 3 image \citep[labelled as F, G, H, and I in][]{bridle94}, in addition to the central feature matching the optical quasar (feature D). Interestingly, the knot with the highest polarization fraction (G with $\sim$30\%) is also the only one that is resolved on the ALMA image. Direct comparison with the 4.9\,GHz map to retrieve a spectral-index map is needed in order to establish whether there is evidence for the location in which the jet goes through realignment as well as its origin \citep[see][]{bridle94,dennett02}. Finally, although hot-spot A is located at the edge of the primary beam, its non-detection, as well as that of hot spot K, is related to the fact that they likely comprise older (less energised) electrons, hence harder to detect at mm wavelengths. The ALMA data were part of the sample analysed by \citet{keenan21}.

SDSS\_J085841.44+140944.8 (3C 212), also part of the \citet{keenan21} sample, is a radio-loud red quasar at a redshift of 1.049 \citep{aldcroft03}, with an X-ray absorption system intrinsic to the quasar itself or its host galaxy  \citep{aldcroft02}. The spatial resolution on the ALMA image is coarser than that of the 3.6\,cm VLA map. Nevertheless, one can still identify the emission associated with the three jet knots in the north-east direction, as well as the two lobe hot spots farther out ($\sim$40 projected kpc). The latter may be related to ongoing supply of energetic electrons and magnetic field to these regions \citep[e.g.,][]{machalski16}, but a multi-frequency analysis is required to confirm this scenario.

SDSS\_J123200.01-022404.7 (4C -02.55) is a quasar at a redshift of 1.05 observed in band 6 \citep{szakacs21}. Comparing with FIRST (1.4\,GHz) and VLASS (2--4\,GHz) imaging, there is a clear morphological difference between mm and cm wavelengths. While ALMA shows four jet knots to the west of the location of the quasar, the VLA image reveals a bright central emission and a fainter source (a jet knot or lobe) to the east of the quasar. This feature falls outside the ALMA field of view. Given the distinct spatial resolutions of each image (i.e. all the structure detectable by ALMA is within a FIRST survey synthesized beam), it is impossible to determine whether this morphology is related to significant jet realignment (i.e. the closest jet component goes from being that in the eastern side to the one in the western side), nevertheless this is the most likely scenario. We finally note that the jet emission detectable by ALMA is co-spatial with the extended emission seen in VLASS Quicklook image, but improved imaging of these data is required to confirm this association.

SDSS\_J135706.53+253724.4 (7C 1354+2552), is a radio-loud quasar at a redshift of 2.007, with two jets that are perpendicular to each other \citep{lonsdale93}. While the north-south direction comprises two isolated lobes (detected only at 1.4\,GHz) whose intermediate point does not coincide with the quasar position, the east-west oriented component is a one-sided jet (to the west of the quasar) and shows remarkable resemblance in 1.4, 5, and 146\,GHz. As reported in \citet{lonsdale93}, the east-west jet is bent (also visible on the ALMA image) and potentially connects with the southern lobe, hinting at a complex nature of the north-south lobes. Molecular gas emission towards the north-east is reported by \citet{vayner21a, vayner21b}. As there is no evidence for cold outflows at the sensitivities of the observations, the molecular gas emission might be related to a merging galaxy.

Finally, SDSS\_J170955.01+223655.7 (4C 22.44) is a radio-loud quasar at a redshift of 1.549. Its radio jets extend out to 7$\arcsec$ along the east–west direction and the two lobes dominate the emission at 3\,GHz as seen in the VLASS Quicklook image \citep{lacy20}. In the ALMA band~4 image one can also see a bright core emission at the location of quasar’s optical coordinates \citep{vayner21a}. The detected source to the north of the quasar does not have a counterpart on the SDSS image and could be a close companion or a chance projection. 

\subsection{Lensed quasars}
\label{sec:lenses}

\begin{figure*}
\captionsetup[subfloat]{labelformat=empty}
  \subfloat[{SDSS\_J081331.28+254503.0 [6]}]{
	\begin{minipage}[c][1\width]{
	   0.3\textwidth}
	   \centering
	   \includegraphics[width=4.5cm]{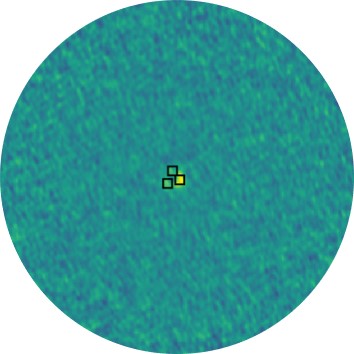}
	\end{minipage}}
 \hfill 	
  \subfloat[{SDSS\_J091127.61+055054.1 [4]}]{
	\begin{minipage}[c][1\width]{
	   0.3\textwidth}
	   \centering
	   \includegraphics[width=4.5cm]{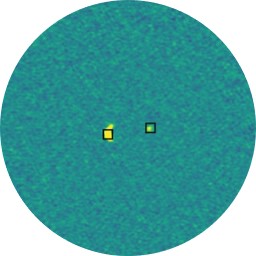}
	\end{minipage}}
 \hfill	
  \subfloat[{SDSS\_J092455.79+021924.9 [7]}]{
	\begin{minipage}[c][1\width]{
	   0.3\textwidth}
	   \centering
	   \includegraphics[width=4.5cm]{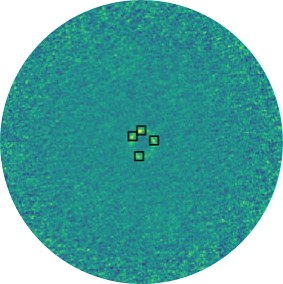}
	\end{minipage}}
	
  \subfloat[{SDSS\_J111816.94+074558.2 [7]}]{
	\begin{minipage}[c][1\width]{
	   0.3\textwidth}
	   \centering
	   \includegraphics[width=4.5cm]{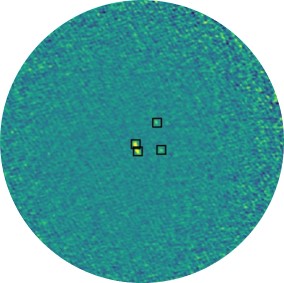}
	\end{minipage}}
 \hfill 	
  \subfloat[{SDSS\_J133018.64+181032.1 [7]}]{
	\begin{minipage}[c][1\width]{
	   0.3\textwidth}
	   \centering
	   \includegraphics[width=4.5cm]{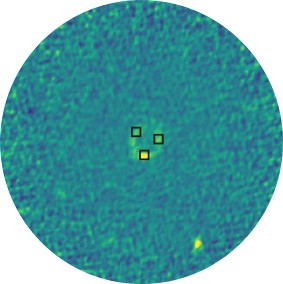}
	\end{minipage}}
 \hfill	
  \subfloat[{SDSS\_J141546.24+112943.4 [7]}]{
	\begin{minipage}[c][1\width]{
	   0.3\textwidth}
	   \centering
	   \includegraphics[width=4.5cm]{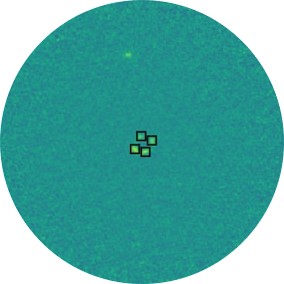}
	\end{minipage}}	
\caption{ALMA cutouts of the lensed SDSS quasars in the ALMA footprint. The corresponding ALMA band is indicated in square brackets next to the name of each object. All cutouts measure 20$^{\prime\prime}$ in diameter while the squares are 1.4\arcsec across.}
\label{fig:lens}
\end{figure*}

Among the quasars in the ALMA footprint, six were already known lensed quasars, with multiple images also visible on the ALMA data, shown in Fig. \ref{fig:lens}. The fluxes of these objects are not included in Table \ref{tab:main}. As noted in Sec. \ref{sec:extraction}, ADMIT does not perform well in the case of complex structures. As a result, the automated extraction does not guarantee reliable fluxes for all the images.  These sub-structures extracted by ADMIT do not always coincide with the real lensed images, or at times a single elongated image (arc) has been broken into more than one sources by the the ADMIT source extraction. Table \ref{tab:lensed} includes all reliable ADMIT extractions, based on visual inspection.

SDSS\_J081331.28+254503.0 (HS 0810+2554), is a radio-quiet quadruply lensed quasar at a redshift of 1.508, observed with ALMA in bands 3 and 4 continuum and CO (3-2) \citep{chartas20, stacey21}. For this object, Fig. \ref{fig:lens} we chose a band 6 image found in the archive, where the multiple images are clearly visible. Observations of this quasar in Chandra and XMM-Newton have revealed highly ionized and relativistic outflows \citep{chartas16}, while VLBI observations revealed a faint jet to dominate the radio emission \citep{hartley19}. 

SDSS\_J091127.61+055054.1 (RXJ0911.4+0551) is also a quadruply lensed quasar at $z$=2.796 observed with ALMA in band 4, targeting also the CO (5-4) transition. Both continuum and CO emission were found to be similarly compact \citep{stacey21}. It was found to have a molecular-gas mass and far-infrared luminosity an order of magnitude lower compared to other quasars. A detailed analysis of the size and morphology of the galaxy \citep{anh13} showed that it was intrinsically a lower mass galaxy rather than an evolved galaxy that had depleted its gas reservoir.

SDSS\_J092455.79+021924.9, is a redshift 1.524 radio-quiet quasar quadruply imaged by a z=0.394 early-type galaxy. The four images are clearly visible in the targeted CO(5-4) line as well as in the band 6 continuum \citep{badole20}. Broad line flux ratios anomalies detected by HST are most likely caused by microlensing effects \citep{morgan06, macleod15}.  

SDSS\_J111816.94+074558.2 (PG1115\_80), is a redshift 1.735 quasar observed with ALMA in band 7, where the four images are clearly visible \citep{stacey21}. Its dust emission was very compact with no apparent clumpy features. While no radio jets have been detected, its radio luminosity indicates signs of star formation. 

SDSS\_J133018.64+181032.1 is a lensed quasar with a quadruple configuration, at $z$=1.393, observed with ALMA in band 7. 

Finally, SDSS\_J141546.24+112943.4 (H1413+117), also known as the Cloverleaf quasar, is a broad absorption line quasar at $z$=2.558, lensed by a galaxy at an unknown redshift. The band 7 continuum clearly shows the four counter-images \citep{stacey21}.

Note that the issue of the missing flux may affect some of the reported fluxes, especially for the more diffuse counterparts. This is a problem inherent to interferometric observations, far from trivial \citep{plunkett23}, and addressing it is well outside the scope of the present work.

\subsection{Quasars with possible nearby submm counterpart}
\label{sec:companions}
During visual inspection 31 unique SDSS quasars with confirmed ALMA counterparts were found to have at least one other ALMA detection within 5$\arcsec$ of the ALMA coordinates. These secondary counterparts could be either close companions in proximity to the primary source or chance associations. Table \ref{tab:multicounterparts} lists the properties of the primary and secondary counterparts in the same format as Table \ref{tab:main}. Cutouts around these objects are displayed in Fig. \ref{fig:companions}. Objects with counterparts already confirmed in the literature are specifically discussed below.

\begin{figure*}
\captionsetup[subfloat]{labelformat=empty}
  \subfloat[{SDSS\_003011.76+004749.9 [7]}]{
	\begin{minipage}[c][1\width]{
	   0.3\textwidth}
	   \centering
	   \includegraphics[width=4.5cm]{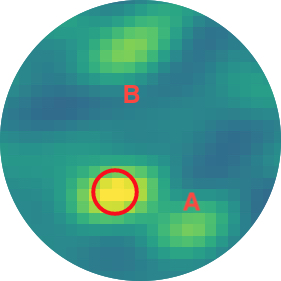}
	\end{minipage}}
 \hfill 	
  \subfloat[{SDSS\_J005021.22+005135.0 [4]}]{
	\begin{minipage}[c][1\width]{
	   0.3\textwidth}
	   \centering
	   \includegraphics[width=4.5cm]{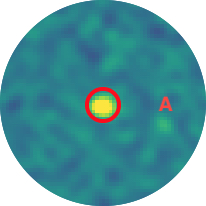}
	\end{minipage}}
 \hfill	
  \subfloat[{SDSS\_J005233.67+014040.8 [4]}]{
	\begin{minipage}[c][1\width]{
	   0.3\textwidth}
	   \centering
	   \includegraphics[width=4.5cm]{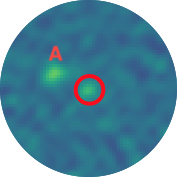}
	\end{minipage}}
	
  \subfloat[{SDSS\_010116.53+020157.3 [4]}]{
	\begin{minipage}[c][1\width]{
	   0.3\textwidth}
	   \centering
	   \includegraphics[width=4.5cm]{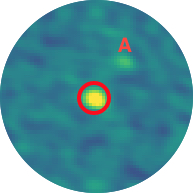}
	\end{minipage}}
 \hfill 	
  \subfloat[{SDSS\_J013825.53-000534.5 [4]}]{
	\begin{minipage}[c][1\width]{
	   0.3\textwidth}
	   \centering
	   \includegraphics[width=4.5cm]{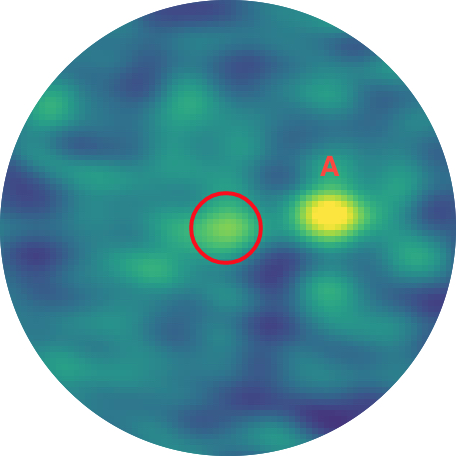}
	\end{minipage}}
 \hfill	
  \subfloat[{SDSS\_J015017.70+002902.4 [7]}]{
	\begin{minipage}[c][1\width]{
	   0.3\textwidth}
	   \centering
	   \includegraphics[width=4.5cm]{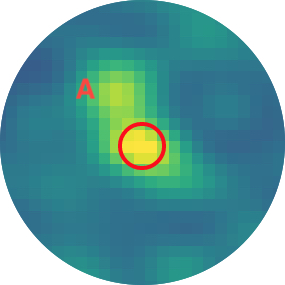}
	\end{minipage}}	
	
  \subfloat[{SDSS\_J021727.65-051502.8 [6]}]{
	\begin{minipage}[c][1\width]{
	   0.3\textwidth}
	   \centering
	   \includegraphics[width=4.5cm]{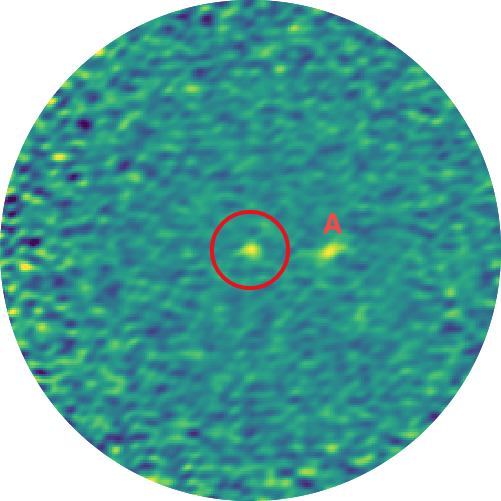}
	\end{minipage}}
 \hfill 	
  \subfloat[{SDSS\_J021757.30-050808.6 [7]}]{
	\begin{minipage}[c][1\width]{
	   0.3\textwidth}
	   \centering
	   \includegraphics[width=4.5cm]{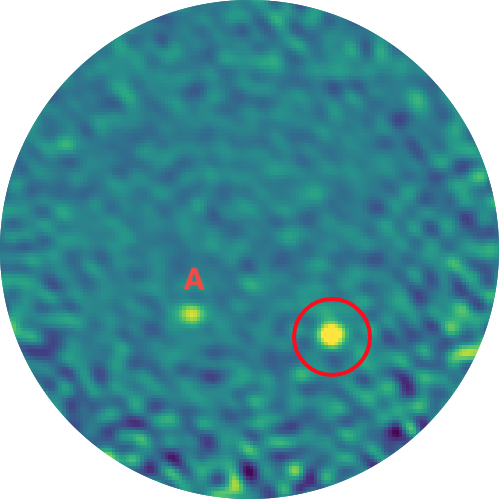}
	\end{minipage}}
 \hfill	
  \subfloat[{SDSS\_J074110.70+311200.2 [5]}]{
	\begin{minipage}[c][1\width]{
	   0.3\textwidth}
	   \centering
	   \includegraphics[width=4.5cm]{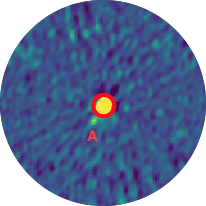}
	\end{minipage}}
	
\caption{Cutouts of SDSS quasars with confirmed ALMA counterparts and secondary ALMA detections at or above a SNR of 3.5 within 5$\arcsec$ of the main quasar counterpart. The corresponding ALMA band is shown in square brackets next to the SDSS name of each object. All cutouts are 20$^{\prime\prime}$. The optical coordinates if the quasars are indicated with a red circle, while the potential submm companions are marked with letters. The circles are 3\arcsec across except for SDSS\_J074110.70+311200.2 (2\arcsec), SDSS\_J141908.17+062834.8 (1.2\arcsec) and SDSS\_J225134.72+184840.0 (0.5\arcsec).}
\label{fig:companions}
\end{figure*}

\begin{figure*}
\ContinuedFloat
\captionsetup[subfloat]{labelformat=empty}
  \subfloat[{SDSS\_J095923.55+022227.2 [6]}]{
	\begin{minipage}[c][1\width]{
	   0.3\textwidth}
	   \centering
	   \includegraphics[width=4.5cm]{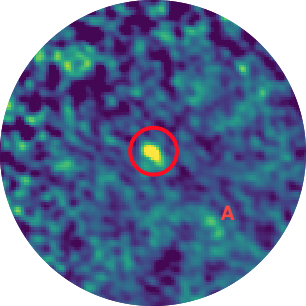}
	\end{minipage}}
 \hfill 	
  \subfloat[{SDSS\_J100038.01+020822.4 [9]}]{
	\begin{minipage}[c][1\width]{
	   0.3\textwidth}
	   \centering
	   \includegraphics[width=4.5cm]{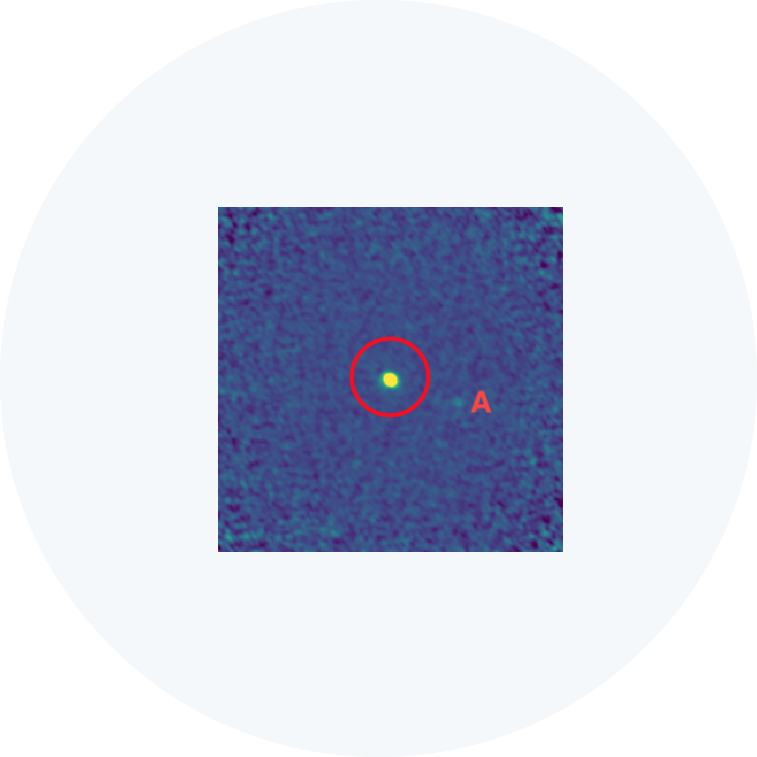}
	\end{minipage}}
 \hfill	
  \subfloat[{SDSS\_J101549.00+002020.0 [7]}]{
	\begin{minipage}[c][1\width]{
	   0.3\textwidth}
	   \centering
	   \includegraphics[width=4.5cm]{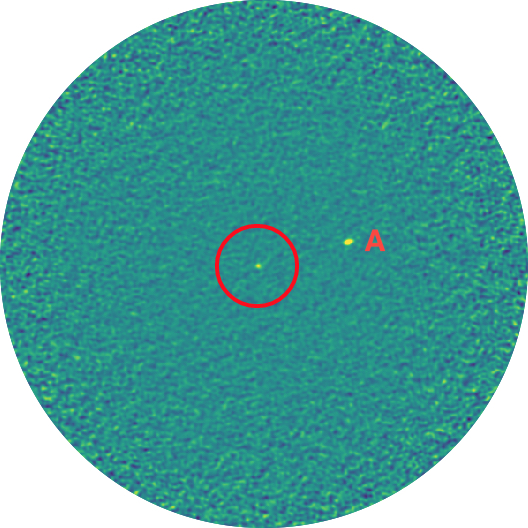}
	\end{minipage}}
	
    \subfloat[{SDSS\_J110045.23+112239.1 [7]}]{
	\begin{minipage}[c][1\width]{
	   0.3\textwidth}
	   \centering
	   \includegraphics[width=4.5cm]{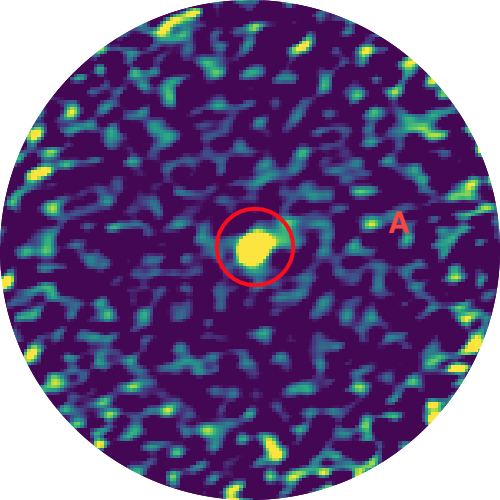}
	\end{minipage}}
 \hfill 	
  \subfloat[{SDSS\_J111200.93+065530.1 [6]}]{
	\begin{minipage}[c][1\width]{
	   0.3\textwidth}
	   \centering
	   \includegraphics[width=4.5cm]{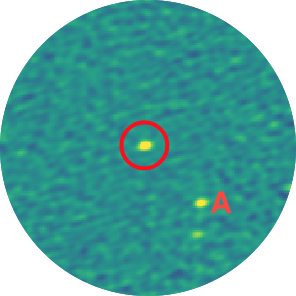}
	\end{minipage}}
 \hfill	
  \subfloat[{SDSS\_J120110.31+211758.4 [8]}]{
	\begin{minipage}[c][1\width]{
	   0.3\textwidth}
	   \centering
	   \includegraphics[width=4.5cm]{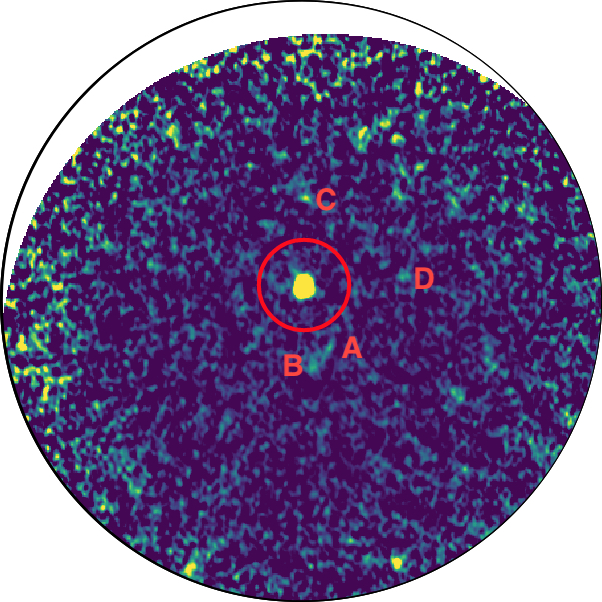}
	\end{minipage}}
	
    \subfloat[{SDSS\_J125353.35+104603.1 [7]}]{
	\begin{minipage}[c][1\width]{
	   0.3\textwidth}
	   \centering
	   \includegraphics[width=4.5cm]{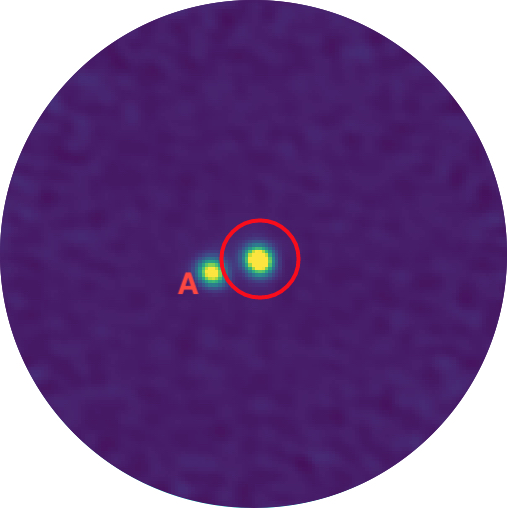}
	\end{minipage}}
 \hfill 	
  \subfloat[{SDSS\_J134131.14+235043.3 [6]}]{
	\begin{minipage}[c][1\width]{
	   0.3\textwidth}
	   \centering
	   \includegraphics[width=4.5cm]{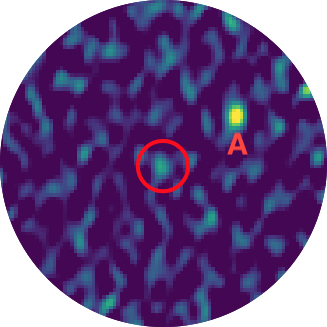}
	\end{minipage}}
 \hfill	
  \subfloat[{SDSS\_J141546.24+112943.4 [7]}]{
	\begin{minipage}[c][1\width]{
	   0.3\textwidth}
	   \centering
	   \includegraphics[width=4.5cm]{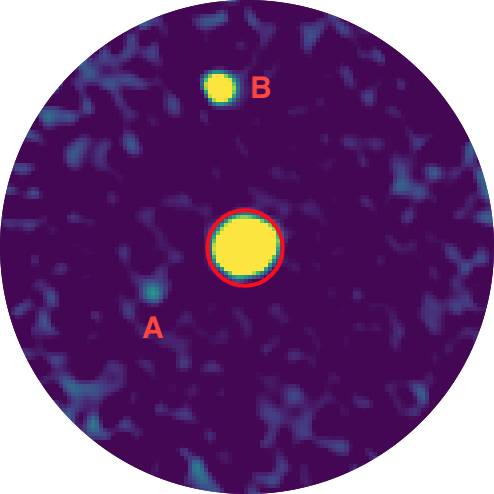}
	\end{minipage}}
\caption{Continued}
\end{figure*}

\begin{figure*}
\ContinuedFloat
\captionsetup[subfloat]{labelformat=empty}
  \subfloat[{SDSS\_J141819.22+044135.0 [6]}]{
	\begin{minipage}[c][1\width]{
	   0.3\textwidth}
	   \centering
	   \includegraphics[width=4.5cm]{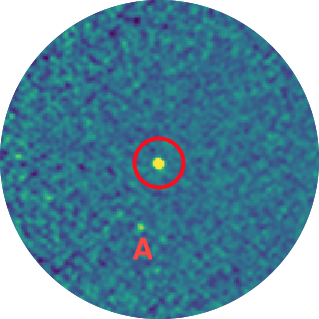}
	\end{minipage}}
 \hfill 	
  \subfloat[{SDSS\_J141908.17+062834.8 [4]}]{
	\begin{minipage}[c][1\width]{
	   0.3\textwidth}
	   \centering
	   \includegraphics[width=4.5cm]{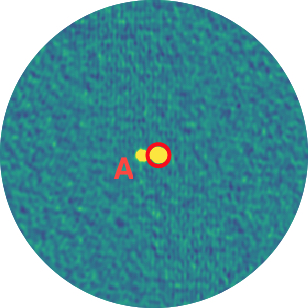}
	\end{minipage}}
 \hfill	
  \subfloat[{SDSS\_J151155.98+040802.9 [7]}]{
	\begin{minipage}[c][1\width]{
	   0.3\textwidth}
	   \centering
	   \includegraphics[width=4.5cm]{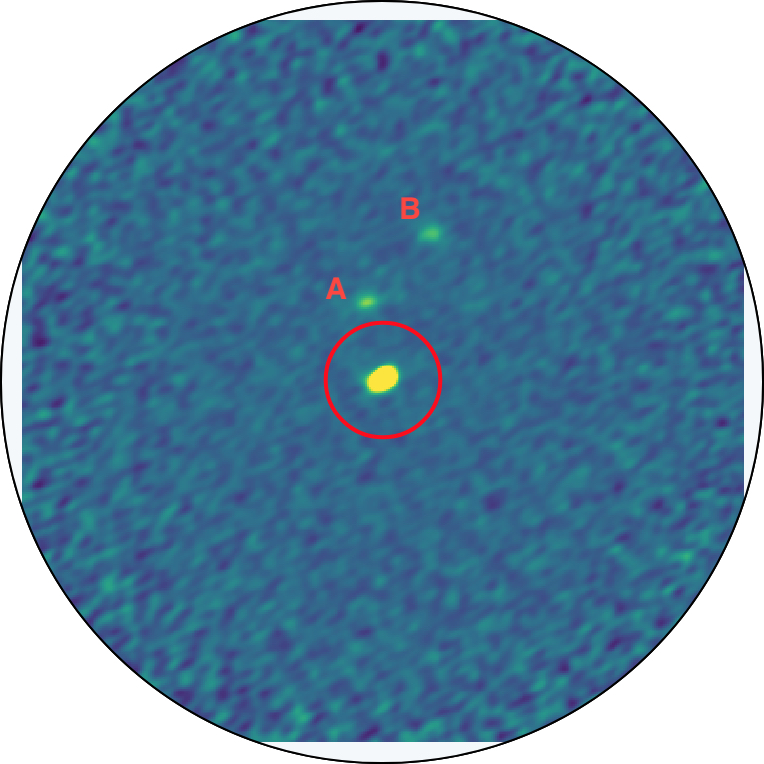}
	\end{minipage}}

  \subfloat[{SDSS\_J154938.71+124509.1 [6]}]{
	\begin{minipage}[c][1\width]{
	   0.3\textwidth}
	   \centering
	   \includegraphics[width=4.5cm]{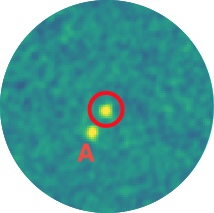}
	\end{minipage}}
 \hfill 	
  \subfloat[{SDSS\_J170955.01+223655.7 [4]}]{
	\begin{minipage}[c][1\width]{
	   0.3\textwidth}
	   \centering
	   \includegraphics[width=4.5cm]{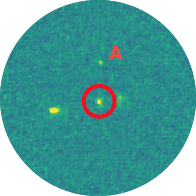}
	\end{minipage}}
 \hfill	
  \subfloat[{SDSS\_J213006.21+001256.6 [6]}]{
	\begin{minipage}[c][1\width]{
	   0.3\textwidth}
	   \centering
	   \includegraphics[width=4.5cm]{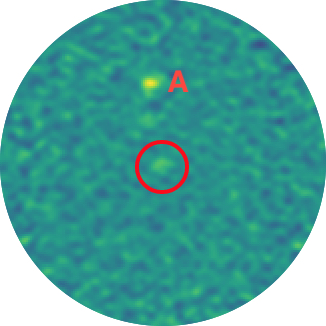}
	\end{minipage}}	

  \subfloat[{SDSS\_J214443.65+002327.9 [6]}]{
	\begin{minipage}[c][1\width]{
	   0.3\textwidth}
	   \centering
	   \includegraphics[width=4.5cm]{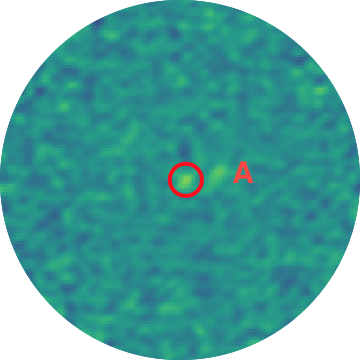}
	\end{minipage}}
 \hfill 	
  \subfloat[{SDSS\_J223619.19+132620.3 [8]}]{
	\begin{minipage}[c][1\width]{
	   0.3\textwidth}
	   \centering
	   \includegraphics[width=4.5cm]{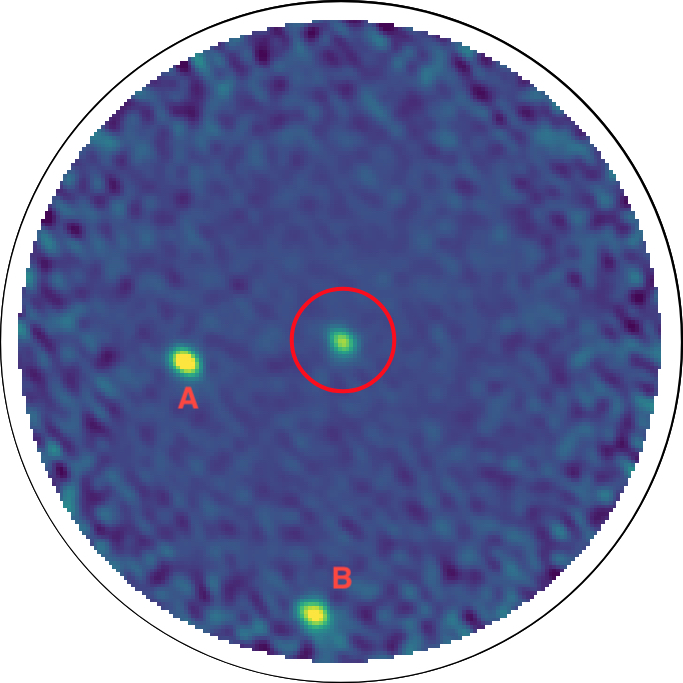}
	\end{minipage}}
 \hfill	
  \subfloat[{SDSS\_J224649.29-004954.3 [7]}]{
	\begin{minipage}[c][1\width]{
	   0.3\textwidth}
	   \centering
	   \includegraphics[width=4.5cm]{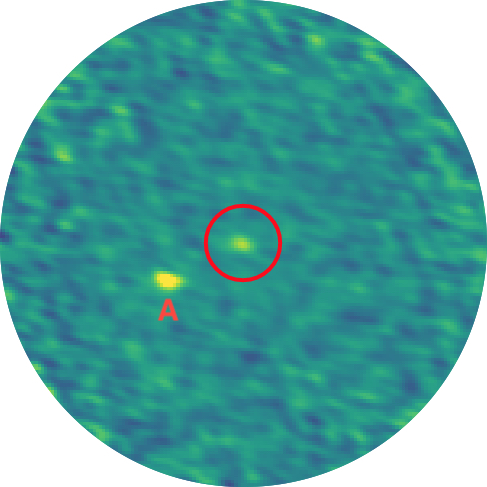}
	\end{minipage}}
\caption{Continued}
\end{figure*}

\begin{figure*}
\ContinuedFloat
\captionsetup[subfloat]{labelformat=empty}
  \subfloat[{SDSS\_J225134.72+184840.0 [7]}]{
	\begin{minipage}[c][1\width]{
	   0.3\textwidth}
	   \centering
	   \includegraphics[width=4.5cm]{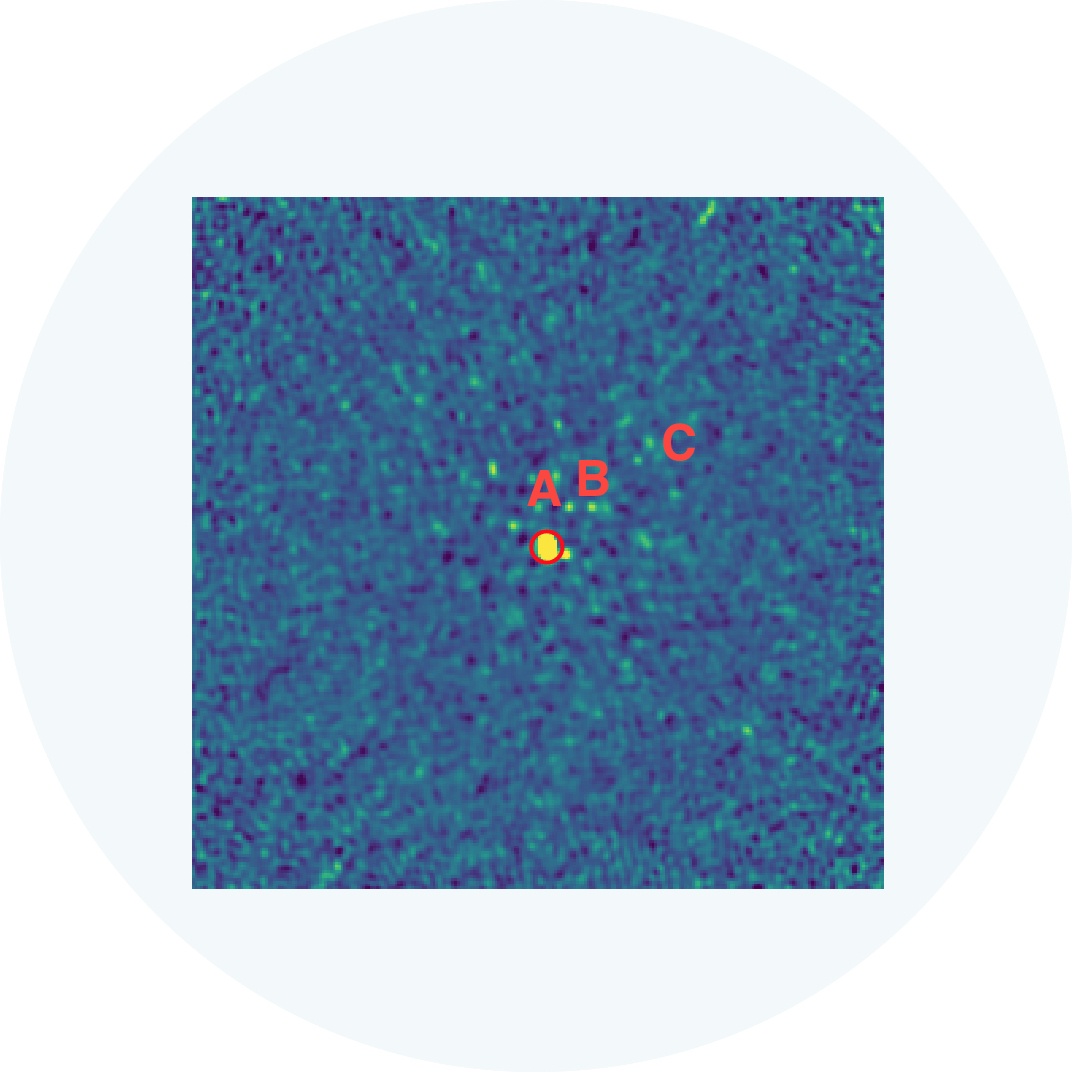}
	\end{minipage}}
 \hfill 	
  \subfloat[{SDSS\_J235637.86+005859.3 [6]}]{
	\begin{minipage}[c][1\width]{
	   0.3\textwidth}
	   \centering
	   \includegraphics[width=4.5cm]{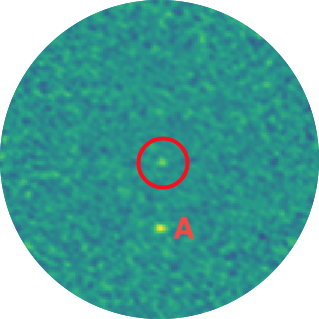}
	\end{minipage}}
 \hfill	
  \subfloat[{SDSS\_J235737.30+003546.2 [6]}]{
	\begin{minipage}[c][1\width]{
	   0.3\textwidth}
	   \centering
	   \includegraphics[width=4.5cm]{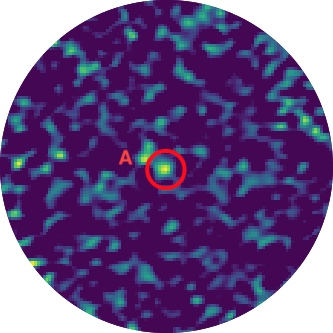}
	\end{minipage}}
	
  \subfloat[{SDSS\_J235944.94+022906.8 [7]}]{
	\begin{minipage}[c][1\width]{
	   0.3\textwidth}
	   \centering
	   \includegraphics[width=4.5cm]{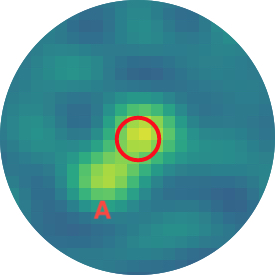}
	\end{minipage}}
\caption{Continued}
\end{figure*}

SDSS\_J003011.76+004749.9, SDSS\_J015017.71+002902.4 and SDSS\_J235944.94+022906.8 were part of the quasars sample observed by \cite{hatzimi18} in the search for multiplicities of the ALMA Compact Array (ACA) counterparts of far-infrared bright quasars. The properties of the counterparts are given in their work, however redshifts of the secondary counterparts were not known and chance associations could not be ruled out. 

SDSS\_J074110.70+311200.2 was found to have two damped Lyman$-\alpha$ (DLA) quasar absorption line systems, one at a redshift of 0.091 and the other at a redshift of 0.221. They were observed and analysed in the optical and infrared wavelength by \cite{turnshek01} and were found to have the lowest redshift of any confirmed DLA absorption-line systems at the time. In Fig. \ref{fig:companions}, only the DLA system at a redshift of 0.2212 is visible and labelled. The other object is not visible, possibly because of its low surface brightness. 

SDSS\_J101549.00+002020.0 was found to have multiple close sources that were detected in band 7 continuum and [CII] line emission less than 4$\arcsec$ away \citep{bischetti18}. Several of these with detected line emission had small velocity shifts with respect to the line emission of the main quasar, and hence were considered close companions. In the cutout shown in Fig. \ref{fig:companions}, only one such companion is visible. This companion was reported by \cite{bischetti18} to have a separation from the main quasar of ~3.5$\arcsec$ and to have a continuum flux of 1252±76 $\mu$Jy. The other detected companions had continuum fluxes $<$400 $\mu$Jy. The low SNR of these detections is the cause of their non-extraction by ADMIT. 

SDSS\_J141908.17+062834.8 was investigated by \cite{vayner17}. They found a kinematic feature located to the south-east of the quasar and offset from by 1.06$\arcsec$, providing evidence of a merging disc system.

SDSS\_J151155.98+040802.9 was observed by \cite{trakhtenbrot17} in band 7 and was confirmed to have two sub-mm companions within 4$\arcsec$ of the main quasar. Both are clearly visible in the continuum image in Fig. \ref{fig:companions}.  
It was concluded that the closer companion was an interacting companion, while the further companion was likely to be a chance association or a minor merger.

SDSS\_J154938.71+124509.1 was observed by \cite{bischetti21} and was found to have two line emitting companion sources within 2.4$\arcsec$ of the quasar. 
Only the further companion was extracted on Fig. \ref{fig:companions} as it is the brighter source. The second fainter companion is located North of the quasar but it is not visible on the image cutout. 

Band 8 observations of SDSS\_J223619.19+132620.3 revealed two nearby submm galaxies \citep{ogura20}. One of them was about 4$\arcsec$ away from the quasar, while the other was about 10$\arcsec$ away. Both sources are also seen in the respective continuum image in Fig. \ref{fig:companions}. 
Since no data in other wavelengths nor spectroscopic redshifts for these objects were available, it was not possible to confirm whether they were physical companions. 

\section{Discussion}
\label{sec:discuss}

This paper describes the first instance of the High-Level Data Products initiative run at the European ALMA Regional Centre. This initiative aims at producing advanced, science-oriented products by combining the ALMA images and cubes publicly available in the ALMA Science Archive. The result of this first instance of HLDP is the creation of a catalogue of 376 submm detections of SDSS DR14 quasars, corresponding to 275 unique quasars, in the ALMA footprint. 

The ALMA observations of the quasars in this work cover, depending on redshift and band, the restframe wavelengths between $\sim$ 190 $\mu$m and $\sim$1.8 mm for the majority of the sample (Fig. \ref{fig:lbolfreq}). The main contributor to the emission at these wavelengths is dust, heated by star formation processes in the quasar hosts, with some contribution from the circumnuclear dust heated by the AGN \citep[e.g.][]{hatzimi10}. Non-thermal (synchrotron) emission from the nucleus, at least in the case of radio-loud quasars, may pitch in to the mm wavelengths, however with single measurements per object it is not possible to distinguish between the thermal and non-thermal components. Nevertheless, the flux measurements provided in the catalogue are relevant to a number of AGN-related studies. 

Most of our current knowledge on the far-infrared (FIR) to mm properties of quasars is based on observations with poor spatial resolution, with e.g. SCUBA-2 or {\it Herschel}/SPIRE (7.9$\arcsec$ at 450 $\mu$m, \citealt{dempsey13} and 18$\arcsec$ at 250 $\mu$m, \citealt{griffin10}, respectively). Because of the poor spatial resolution, identifying counterparts of FIR sources at other wavelengths has long been an issue \citep[e.g.][]{wang14, hurley17}. This in turns means that the spectral energy distributions (SEDs) of quasars at wavelengths beyond a few tens of micron are often poorly defined, consequently affecting properties derived from those SEDs, such as SFRs in quasar hosts, dust temperatures etc.

Thanks to their high spatial resolution, ALMA observations of FIR-bright quasars can confirm the source of the FIR/submm emission \citep[e.g.][]{hatzimi18}, help evaluate the accuracy of the FIR fluxes and add data points to the FIR SEDs. This, in turn, results in a more accurate estimate of the total FIR emission and, subsequently, to the calibration of the FIR-derived SFRs. In case of multiple submm counterparts to the FIR sources, the FIR emission can be redistributed among those counterparts and the SEDs can be improved, with a subsequent improvement of the SFR estimates. 

And this is only a tiny sample of science cases that can be addressed with data sets like the one produced in this work. Upcoming instances of ALMA HLDP and HLDP-like initiatives will cater for a large palette of science cases. The potential for the creation of value-added, science-ready data products is large and it is increasing as more data are added to the ASA every day, continuously providing new possibilities for original, ground-breaking research.

\section*{Acknowledgements}

AW would like to thank the IAC and ESO for their hospitality for a three- and a four-month visiting periods, respectively, during which part of this work was pursued, and Nanyang Technological University and the CN Yang Scholarship Programme for funding his expenses for this project. EH acknowledges support from the Fundaci\'on Jes\'us Serra and the Instituto de Astrof\'isica de Canarias (IAC) under the Visiting Researcher Programme 2020-2022 agreed between both institutions. AB acknowledges the support of the EU-ARC.CZ Large Research Infrastructure grant project LM2023059 of the Ministry of Education, Youth and Sports of the Czech Republic and the Czech Science Foundation project No.19-05599Y. IPF and FP acknowledge support from the Spanish State Research Agency (AEI) under grant numbers PID2019-105552RB-C43.
This work makes use of TOPCAT, ``TOPCAT \& STIL: Starlink Table/VOTable Processing Software'', M. B. Taylor, in Astronomical Data Analysis Software and Systems XIV, eds. P Shopbell et al., ASP Conf. Ser. 347, p. 29, 2005; Astropy (Astropy Collaboration et al. 2013, 2018); Numpy (Harris et al. 2020 https://numpy.org/citing-numpy/); Matplotlib (Hunter 2007). 
This paper makes use of the following ALMA data: ADS/JAO.ALMA\#2012.1.00175.S, 
ADS/JAO.ALMA\#2012.1.00611.S,
ADS/JAO.ALMA\#2013.1.00535.S, 
ADS/JAO.ALMA\#2013.1.00576.S, 
ADS/JAO.ALMA\#2013.1.01153.S, 
ADS/JAO.ALMA\#2013.1.01178.S,
ADS/JAO.ALMA\#2013.1.01342.S,
ADS/JAO.ALMA\#2013.1.01359.S,
ADS/JAO.ALMA\#2015.1.00137.S,
ADS/JAO.ALMA\#2015.1.00260.S,
ADS/JAO.ALMA\#2015.1.00305.S,
ADS/JAO.ALMA\#2015.1.00329.S,
ADS/JAO.ALMA\#2015.1.00582.S,
ADS/JAO.ALMA\#2015.1.00754.S,
ADS/JAO.ALMA\#2015.1.00856.S,
ADS/JAO.ALMA\#2015.1.00932.S,
ADS/JAO.ALMA\#2015.1.01034.S,
ADS/JAO.ALMA\#2015.1.01090.S,
ADS/JAO.ALMA\#2015.1.01147.S,
ADS/JAO.ALMA\#2015.1.01148.S,
ADS/JAO.ALMA\#2015.1.01309.S,
ADS/JAO.ALMA\#2015.1.01362.S,
ADS/JAO.ALMA\#2015.1.01366.S,
ADS/JAO.ALMA\#2015.1.01452.S,
ADS/JAO.ALMA\#2015.1.01469.S,
ADS/JAO.ALMA\#2015.1.01564.S,
ADS/JAO.ALMA\#2016.1.00434.S,
ADS/JAO.ALMA\#2016.1.00463.S,
ADS/JAO.ALMA\#2016.1.00569.S,
ADS/JAO.ALMA\#2016.1.00718.S,
ADS/JAO.ALMA\#2016.1.00735.S,
ADS/JAO.ALMA\#2016.1.00798.S,
ADS/JAO.ALMA\#2016.1.00864.S,
ADS/JAO.ALMA\#2016.1.01326.S,
ADS/JAO.ALMA\#2016.1.01481.S,
ADS/JAO.ALMA\#2016.1.01515.S,
ADS/JAO.ALMA\#2016.2.00060.S,
ADS/JAO.ALMA\#2017.1.00102.S,
ADS/JAO.ALMA\#2017.1.00297.S,
ADS/JAO.ALMA\#2017.1.00345.S,
ADS/JAO.ALMA\#2017.1.00358.S,
ADS/JAO.ALMA\#2017.1.00478.S,
ADS/JAO.ALMA\#2017.1.00560.S,
ADS/JAO.ALMA\#2017.1.00571.S,
ADS/JAO.ALMA\#2017.1.00963.S,
ADS/JAO.ALMA\#2017.1.01027.S,
ADS/JAO.ALMA\#2017.1.01052.S,
ADS/JAO.ALMA\#2017.1.01081.S,
ADS/JAO.ALMA\#2017.1.01232.S,
ADS/JAO.ALMA\#2017.1.01324.S,
ADS/JAO.ALMA\#2017.1.01332.S,
ADS/JAO.ALMA\#2017.1.01368.S,
ADS/JAO.ALMA\#2017.1.01492.S,
ADS/JAO.ALMA\#2017.1.01527.S,
ADS/JAO.ALMA\#2017.1.01559.S,
ADS/JAO.ALMA\#2017.1.01572.S,
ADS/JAO.ALMA\#2017.1.01676.S,
ADS/JAO.ALMA\#2018.1.00526.S,
ADS/JAO.ALMA\#2018.1.00583.S,
ADS/JAO.ALMA\#2018.1.00585.S,
ADS/JAO.ALMA\#2018.1.00681.S,
ADS/JAO.ALMA\#2018.1.00699.S,
ADS/JAO.ALMA\#2018.1.00828.S,
ADS/JAO.ALMA\#2018.1.00859.S,
ADS/JAO.ALMA\#2018.1.00932.S,
ADS/JAO.ALMA\#2018.1.00992.S,
ADS/JAO.ALMA\#2018.1.01008.S,
ADS/JAO.ALMA\#2018.1.01044.S,
ADS/JAO.ALMA\#2018.1.01368.S,
ADS/JAO.ALMA\#2018.1.01447.S,
ADS/JAO.ALMA\#2018.1.01464.S,
ADS/JAO.ALMA\#2018.1.01476.S,
ADS/JAO.ALMA\#2018.1.01575.S,
ADS/JAO.ALMA\#2018.1.01591.S,
ADS/JAO.ALMA\#2018.1.01784.S,
ADS/JAO.ALMA\#2018.1.01806.S,
ADS/JAO.ALMA\#2018.1.01817.S,
ADS/JAO.ALMA\#2018.1.01830.S,
ADS/JAO.ALMA\#2019.1.00403.S,
ADS/JAO.ALMA\#2019.1.00411.S,
ADS/JAO.ALMA\#2019.1.00477.S
ADS/JAO.ALMA\#2019.1.00840.S,
ADS/JAO.ALMA\#2019.1.00883.S,
ADS/JAO.ALMA\#2019.1.00946.S,
ADS/JAO.ALMA\#2019.1.00948.S,
ADS/JAO.ALMA\#2019.1.00959.S,
ADS/JAO.ALMA\#2019.1.01022.S,
ADS/JAO.ALMA\#2019.1.01070.S,
ADS/JAO.ALMA\#2019.1.01251.S,
ADS/JAO.ALMA\#2019.1.01422.S,
ADS/JAO.ALMA\#2019.1.01514.S,
ADS/JAO.ALMA\#2019.1.01587.S,
ADS/JAO.ALMA\#2019.1.01709.S,
ADS/JAO.ALMA\#2019.1.01802.S,
ADS/JAO.ALMA\#2019.2.00034.S,
ADS/JAO.ALMA\#2019.2.00085.S. ALMA is a partnership of ESO (representing its member states), NSF (USA) and NINS (Japan), together with NRC (Canada), MOST and ASIAA (Taiwan), and KASI (Republic of Korea), in cooperation with the Republic of Chile. The Joint ALMA Observatory is operated by ESO, AUI/NRAO and NAOJ.

\section*{Data Availability}

The full versions of the catalogues produced in this paper are available as online supplementary material. The ALMA data used to produce the catalogues are all available in the ASA. 



\bibliographystyle{mnras}
\bibliography{bibliography.bib} 




\appendix

\section{Querying the ALMA archive using astroquery}
\label{sec:astroquery}

The astroquery commands below allow for the retrieval of the MOUS IDs that include SDSS quasar positions.\\

\noindent
\begin{lstlisting}
import astropy.units as u
from astropy import coordinates
from astropy.io import ascii
import pyvo
import pandas as pd

service = pyvo.dal.TAPService( "https://almascience.eso.org/tap")

# Define input dataset
fileDir = '/your/input/directory/'
fileName = 'input_file.csv'

# Define output dataset
outfile = 'output_file.csv'

# Read in the input dataset
reader = ascii.read(fileDir+fileName, format='basic', delimiter=',')

# Selecting relevant parameters from the input dataset
name = reader['SDSS']  # SDSS name
coord = coordinates.SkyCoord(ra=reader['RAJ2000'],
                             dec=reader['DEJ2000'],
                             frame='icrs', unit=(u.deg, u.deg))

# Function to query ALMA archive using coordinates and save as Pandas DataFrame
# Sources observed as or in the field of view of TARGET observations


def query_coord(service, ra, dec, radius):
    """ra, dec and radius in decimal degrees """

    query = f"""
            SELECT *
            FROM ivoa.obscore
            WHERE INTERSECTS(CIRCLE('ICRS',{ra},{dec},
                  {radius}),
                  s_region)=1
            and scan_intent like '%TARGET%'"""

    return service.search(query).to_table().to_pandas()

# Sources observed as or in the field of view of calibrator observations


def query_coord_cal(service, ra, dec, radius):
    """ra, dec and radius in decimal degrees """

    query = f"""
            SELECT *
            FROM ivoa.obscore
            WHERE INTERSECTS(CIRCLE('ICRS',{ra},{dec},
                  {radius}),
                  s_region)=1
            and scan_intent not like '%TARGET%'"""

    return service.search(query).to_table().to_pandas()


# Initiate importat ALMA parameter columns to save
SDSS_ind = []
ra = []
dec = []
propIDs = []
SBs = []
bands = []
MOUSs = []
URLs = []
keys = []
sci_cats = ['Active galaxies', 'Cosmology', 
            'Galaxy evolution', 'Local Universe']
non_AGN = []

for j in range(0, len(ind), 100):
    # query 100 objects at a time to not overwhelm the servers.
    for i, CQ in zip(ind[j:j+100], coord[j:j+100]):
    AQ = query_coord(service, CQ.ra.degree, CQ.dec.degree, 0.0084)
    if len(AQ) > 0 and str(AQ['scientific_category'][0]) in sci_cats:
        SDSS_ind.extend([i] * len(AQ))
        ra.extend([CQ.ra.degree] * len(AQ))
        dec.extend([CQ.dec.degree] * len(AQ))
        propIDs.extend(AQ['proposal_id'])
        SBs.extend(AQ['schedblock_name'])
        bands.extend(AQ['band_list'])
        MOUSs.extend(AQ['member_ous_uid'])
        keys.extend(AQ['scientific_category'])
        URLs.extend(AQ['access_url'])
    if len(AQ) > 0 and str(AQ['scientific_category'][0]) not in sci_cats:
        non_AGN.extend([i] * len(AQ))

# Create a pandas DataFrame from the selected columns
T = pd.DataFrame({'SDSS_index': SDSS_ind,
                  'SDSS_RA': ra,
                  'SDSS_DEC': dec,
                  'proposal_id': propIDs,
                  'schedblock_name': SBs,
                  'band_list': bands,
                  'member_ous_uid': MOUSs,
                  'access_url': URLs,
                  'scientific_category': keys}).drop_duplicates()

# Save the DataFrame
T.to_csv(outfile, index=False)
pd.DataFrame({'non_AGN_index': non_AGN}).to_csv('non_AGN_index.csv', index=False)
\end{lstlisting}

\section{catalogues}
\label{sec:catalogues}

The tables discussed in the paper are presented in his appendix. 

Table \ref{tab:main} contains the first 20 entries of the full catalogue of submm detections of DR14Q quasars in the ALMA footprint. Table \ref{tab:multicounterparts} shows the first 21 lines of the list of DR14Q quasars in the ALMA footprint with more than one submm counterparts extracted within 5$\arcsec$ from the optical coordinates of the quasar. There are cases for which more counterparts are visible on the images than the ones identified by the automated extraction method. The number in parenthesis next to the SDSS name indicates, for each object, how many of those counterparts without flux measurements are visible on the image. Table \ref{tab:lensed} lists the flux of the various visually confirmed sub-structures identified on the images of and around the lensed quasars discussed in Sec. \ref{sec:lenses}. Table \ref{tab:NonDetected} lists the first 20 of the DR14Q quasars in the ALMA footprint for which no submm counterpart was extracted. Finally, Table \ref{tab:pbc} contains the four quasars with visible submm counterparts for which no primary-beam correction curve was available and for which the automated extraction did not return any information. For a description of the columns in the various catalogues see the table captions.

\begin{landscape}

\begin{table}
\caption{The first 20 lines in the main catalogue of DR14Q quasars with ALMA detections. In order, the columns are: SDSS name, RA and Dec; redshift; ALMA RA and Dec; separation (Sep.) between the SDSS and ALMA coordinates in $\arcsec$; the ALMA band and array used for the observations; the angular resolution (Ang. res.) of the ALMA image in $\arcsec$; the continuum sensitivity (Sens.) of the ALMA image in mJy;  
the peak and integrated fluxes (S$_{\rm peak}$ and S$_{\rm int}$) and corresponding SNRs; the major and minor axes of the extracted 2D Gaussian ($\alpha_{min}$ and $\alpha_{max}$, respectively) and the position angle of the ellipse (PA); the major and minor axes of the ALMA beam ($\alpha^b_{min}$ and $\alpha^b_{max}$, respectively) and associated  position angle (PA$^b$); the RMS of the image; the date of observation (the date is that of the first execution in case of multiple executions); the ALMA project code and MOUS ID; and an indicator as to whether the object is part of ALMACAL.}
\adjustbox{width=1.33\textwidth}{
\centering
\label{tab:main}
\begin{tabular}{lcccccccccclclllccccrrcclc}
    \hline
    \multicolumn{4}{c}{SDSS} &
    \multicolumn{2}{c}{ALMA} &
    Sep. &
    Band &
    Array &
    Ang. res. &
    Sens. &
    S$_{\rm peak}$ &
    S$_{\rm peak}$  &
    S$_{\rm int}$ &
    S$_{\rm int}$  &
    $\alpha_{max}$ &
    $\alpha_{min}$ &
    PA &
    $\alpha_{max}^{b}$ &
    $\alpha_{min}^{b}$ &
    PA$^{b}$ &
    RMS &
    Obs. date &
    Project code &
    MOUS ID &
    ALMACAL \\
    Name &
    RA &
    Dec &
    $z$ &
    RA &
    Dec &
    ($\arcsec$) &
     & 
     & 
    ($\arcsec$) &
    (mJy) &
    (mJy) &
    SNR &
    (mJy) &
    SNR &
    ($\arcsec$) &
    ($\arcsec$) &
    ($^{\circ}$) &
    ($\arcsec$) &
    ($\arcsec$) &
    ($^{\circ}$) &
    (mJy) &
     & 
     & 
     & 
     \\
    \hline
  000413.63-085529.5 & 00:04:13.64 & -08:55:29.6 & 2.435 & 00:04:13.64 & -08:55:29.6 & 0.05 & 6 & 12m & 0.70 & 0.020 & 0.4 & 13.0 & 0.4 & 12.7 & 0.7 & 0.6 & 67 & 0.7 & 0.5 & 85 & 0.03 & 2018-12-27 & 2018.1.00583.S & uid://A001/X1374/X7e & false\\
  000610.67+121501.2 & 00:06:10.68 & +12:15:01.3 & 2.309 & 00:06:10.68 & +12:15:01.4 & 0.15 & 6 & 12m & 0.78 & 0.013 & 0.8 & 24.5 & 1.0 & 25.1 & 1.0 & 0.6 & 69 & 0.9 & 0.5 & 66 & 0.03 & 2018-08-24 & 2017.1.00478.S & uid://A001/X12d1/X19a & false\\
  000746.92+001543.0 & 00:07:46.93 & +00:15:43.0 & 2.479 & 00:07:46.93 & +00:15:43.0 & 0.14 & 7 & 7m & 3.44 & 0.963 & 5.5 & 11.2 & 6.4 & 15.4 & 3.7 & 3.0 & 75 & 4.3 & 3.1 & 88 & 0.49 & 2017-07-19 & 2016.2.00060.S & uid://A001/X124a/X16e & false\\
  001121.86-000918.6 & 00:11:21.87 & -00:09:18.6 & 3.009 & 00:11:21.98 & -00:09:19.6 & 2.00 & 7 & 7m & 3.44 & 0.961 & 7.5 & 14.6 & 9.9 & 17.5 & 4.7 & 3.2 & 109 & 4.3 & 3.1 & 88 & 0.51 & 2017-07-19 & 2016.2.00060.S & uid://A001/X124a/X16e & false\\
  001401.09-010607.1 & 00:14:01.09 & -01:06:07.2 & 2.103 & 00:14:01.11 & -01:06:07.0 & 0.35 & 7 & 7m & 3.44 & 0.963 & 6.4 & 15.7 & 6.5 & 17.0 & 4.4 & 2.9 & 89 & 4.3 & 3.1 & 87 & 0.41 & 2017-07-19 & 2016.2.00060.S & uid://A001/X124a/X16e & false\\
  002025.22+154054.7 & 00:20:25.22 & +15:40:54.7 & 2.019 & 00:20:25.22 & +15:40:54.8 & 0.06 & 3 & 12m & 0.43 & 0.033 & 0.6 & 6.5 & 0.8 & 6.8 & 0.6 & 0.4 & 177 & 0.5 & 0.3 & 0 & 0.09 & 2016-09-08 & 2015.1.00932.S & uid://A001/X2f6/X209 & false\\
   &  &  &  & 00:20:25.23 & +15:40:54.8 & 0.12 & 6 & 12m & 0.48 & 0.038 & 0.3 & 3.9 & 0.3 & 4.2 & 0.5 & 0.4 & 97 & 0.6 & 0.4 & 33 & 0.08 & 2019-05-04 & 2018.1.01817.S & uid://A001/X137b/X2e & false\\
   &  &  &  & 00:20:25.22 & +15:40:54.8 & 0.09 & 6 & 12m & 0.43 & 0.065 & 0.3 & 4.0 & 0.4 & 4.0 & 0.8 & 0.4 & 17 & 0.6 & 0.5 & 8 & 0.08 & 2016-12-02 & 2016.1.01481.S & uid://A001/X885/Xb4 & false\\
   &  &  &  & 00:20:25.22 & +15:40:54.8 & 0.08 & 6 & 12m & 0.43 & 0.029 & 0.2 & 6.0 & 0.2 & 6.1 & 0.2 & 0.1 & 40 & 0.2 & 0.1 & 15 & 0.04 & 2016-10-01 & 2016.1.01481.S & uid://A001/X885/Xb2 & false\\
   &  &  &  & 00:20:25.22 & +15:40:54.8 & 0.04 & 6 & 12m & 0.43 & 0.032 & 0.3 & 4.8 & 0.4 & 5.1 & 0.4 & 0.3 & 21 & 0.4 & 0.3 & 40 & 0.07 & 2016-07-28 & 2015.1.00932.S & uid://A001/X2f6/X20d & false\\
  003011.78+004749.9 & 00:30:11.78 & +00:47:50.0 & 3.118 & 00:30:11.77 & +00:47:50.1 & 0.18 & 7 & 7m & 3.32 & 0.398 & 6.4 & 11.3 & 6.9 & 11.9 & 4.8 & 3.0 & 90 & 4.9 & 2.9 & -77 & 0.57 & 2017-07-04 & 2016.2.00060.S & uid://A001/X124a/X174 & false\\
  003126.79+150739.5 & 00:31:26.80 & +15:07:39.5 & 4.290 & 00:31:26.80 & +15:07:39.5 & 0.05 & 7 & 12m & 0.20 & 0.024 & 1.6 & 55.7 & 1.8 & 58.3 & 0.2 & 0.2 & 44 & 0.2 & 0.2 & 39 & 0.03 & 2016-07-27 & 2015.1.01469.S & uid://A001/X5a4/X29e & false\\
  003820.53-020740.4 & 00:38:20.53 & -02:07:40.4 & 0.220 & 00:38:20.53 & -02:07:40.6 & 0.14 & 3 & 12m & 1.75 & 0.021 & 176.0 & 334.0 & 240.0 & 345.0 & 2.7 & 2.2 & 61 & 2.4 & 1.9 & 64 & 0.53 & 2019-11-03 & 2019.1.01022.S & uid://A001/X1465/X1d6b & false\\
  004017.42+170619.7 & 00:40:17.43 & +17:06:19.8 & 3.914 & 00:40:17.43 & +17:06:19.8 & 0.11 & 3 & 12m & 0.77 & 0.014 & 0.1 & 8.4 & 0.2 & 8.4 & 1.1 & 1.1 & 169 & 1.0 & 0.8 & 36 & 0.01 & 2020-03-17 & 2019.1.00411.S & uid://A001/X1465/X3053 & false\\
  004054.65-091526.8 & 00:40:54.65 & -09:15:26.8 & 4.976 & 00:40:54.66 & -09:15:26.8 & 0.08 & 7 & 12m & 0.75 & 0.029 & 4.8 & 196.0 & 5.5 & 204.0 & 0.8 & 0.7 & 91 & 0.8 & 0.6 & -86 & 0.02 & 2018-05-16 & 2016.1.00569.S & uid://A001/X879/X201 & false\\
  004219.74-102009.5 & 00:42:19.75 & -10:20:09.5 & 3.880 & 00:42:19.75 & -10:20:09.6 & 0.08 & 3 & 12m & 1.05 & 0.009 & 0.0 & 3.6 & 0.1 & 4.4 & 1.5 & 0.8 & 125 & 1.1 & 0.9 & -76 & 0.01 & 2019-10-19 & 2019.1.00411.S & uid://A001/X1465/X3057 & false\\
  004440.49+010306.4 & 00:44:40.50 & +01:03:06.4 & 3.288 & 00:44:40.49 & +01:03:06.7 & 0.30 & 7 & 7m & 3.32 & 0.398 & 8.3 & 20.1 & 8.1 & 21.5 & 4.6 & 2.7 & 99 & 4.8 & 2.9 & -78 & 0.41 & 2017-07-04 & 2016.2.00060.S & uid://A001/X124a/X174 & false\\
  004730.35+042304.7 & 00:47:30.36 & +04:23:04.7 & 3.864 & 00:47:30.36 & +04:23:04.7 & 0.06 & 3 & 12m & 1.04 & 0.009 & 0.1 & 5.6 & 0.0 & 6.1 & 0.9 & 0.9 & 87 & 1.1 & 1.0 & -61 & 0.01 & 2019-10-14 & 2019.1.00411.S & uid://A001/X1465/X305b & false\\
  005021.22+005135.0 & 00:50:21.23 & +00:51:35.1 & 2.247 & 00:50:21.23 & +00:51:35.1 & 0.07 & 4 & 12m & 1.80 & 0.013 & 0.3 & 20.6 & 0.3 & 21.5 & 1.9 & 1.5 & 88 & 1.8 & 1.6 & 79 & 0.01 & 2019-11-22 & 2019.1.01251.S & uid://A001/X1465/X1315 & false\\
  005130.48+004150.0 & 00:51:30.49 & +00:41:49.9 & 1.189 & 00:51:30.49 & +00:41:50.0 & 0.09 & 4 & 12m & 0.24 & 0.015 & 1.5 & 41.6 & 1.6 & 41.9 & 0.3 & 0.2 & 104 & 0.3 & 0.2 & -74 & 0.04 & 2017-12-28 & 2017.1.01559.S & uid://A001/X1296/Xab1 & false\\
\hline
\end{tabular}}
\end{table}
\end{landscape}

\clearpage

\begin{landscape}

\begin{table}
\caption{The first 21 lines in the catalogue of SDSS quasars with confirmed ALMA counterparts that have possible close companions within 5\arcsec. The primary ALMA counterpart for each object (i.e. the ALMA source with the smallest separation from the SDSS coordinates) is not included in the table, as they are part of Table \ref{tab:main}. The catalogue contains the same columns as in Table \ref{tab:main}. The number in parenthesis next to the SDSS name indicates the number of secondary counterparts visible on the images but with no flux extraction from the automated process.}
\adjustbox{width=1.33\textwidth}{
\centering
\label{tab:multicounterparts}
\begin{tabular}{lcccccccccclllllclccrrcclc}
    \hline
    \multicolumn{4}{c}{SDSS} &
    \multicolumn{2}{c}{ALMA} &
    Sep. &
    Band &
    Array &
    Ang. res. &
    Sens. &
    S$_{\rm peak}$ &
    S$_{\rm peak}$  &
    S$_{\rm int}$ &
    S$_{\rm int}$  &
    $\alpha_{max}$ &
    $\alpha_{min}$ &
    PA &
    $\alpha_{max}^{b}$ &
    $\alpha_{min}^{b}$ &
    PA$^{b}$ &
    RMS &
    Obs. date &
    Project code &
    MOUS ID &
    ALMACAL \\
    Name &
    RA &
    Dec &
    $z$ &
    RA &
    Dec &
    ($\arcsec$) &
     & 
     & 
    ($\arcsec$) &
    (mJy) &
    (mJy) &
    SNR &
    (mJy) &
    SNR &
    ($\arcsec$) &
    ($\arcsec$) &
    ($^{\circ}$) &
    ($\arcsec$) &
    ($\arcsec$) &
    ($^{\circ}$) &
    (mJy) &
     & 
     & 
     & 
     \\
    \hline
  003011.78+004749.9 (2) & 00:30:11.78 & +00:47:50.0 & 3.118 & 00:30:11.78 & +00:47:50.1 & 0.18 & 7 & 7m & 3.32 & 0.398 & 6.4 & 11.3 & 6.9 & 11.9 & 4.8 & 3.0 & 90 & 4.9 & 2.9 & -77 & 0.57 & 2017-07-04 & 2016.2.00060.S & uid://A001/X124a/X174 & false\\
  005021.22+005135.0 (1) & 00:50:21.23 & +00:51:35.1 & 2.247 & 00:50:21.23 & +00:51:35.1 & 0.07 & 4 & 12m & 1.80 & 0.013 & 0.3 & 20.6 & 0.3 & 21.5 & 1.9 & 1.5 & 88 & 1.8 & 1.6 & 79 & 0.01 & 2019-11-22 & 2019.1.01251.S & uid://A001/X1465/X1315 & false\\
  005233.67+014040.8 (1) & 00:52:33.67 & +01:40:40.9 & 2.301 & 00:52:33.67 & +01:40:40.8 & 0.14 & 4 & 12m & 1.86 & 0.013 & 0.1 & 9.2 & 0.1 & 8.9 & 2.0 & 1.7 & 135 & 1.9 & 1.6 & -75 & 0.01 & 2019-11-23 & 2019.1.01251.S & uid://A001/X1465/X131b & false\\
  010116.53+020157.3 (1) & 01:01:16.53 & +02:01:57.4 & 2.443 & 01:01:16.53 & +02:01:57.6 & 0.21 & 4 & 12m & 1.72 & 0.013 & 0.2 & 16.7 & 0.2 & 17.9 & 2.1 & 1.5 & 67 & 2.0 & 1.5 & 71 & 0.01 & 2019-11-16 & 2019.1.01251.S & uid://A001/X1465/X1321 & false\\
  013825.53-000534.5 & 01:38:25.54 & -00:05:34.6 & 1.340 & 01:38:25.53 & -00:05:34.3 & 0.26 & 4 & 12m & 2.06 & 0.013 & 0.1 & 3.9 & 0.1 & 4.1 & 3.3 & 2.0 & 104 & 2.8 & 2.0 & -88 & 0.02 & 2015-12-29 & 2015.1.01034.S & uid://A001/X2d8/X8b & false\\
   &  &  &  & 01:38:25.23 & -00:05:33.7 (A) & 4.77 & 4 & 12m & 2.06 & 0.013 & 0.1 & 7.0 & 0.1 & 7.2 & 2.7 & 2.0 & 91 & 2.8 & 2.0 & -88 & 0.02 & 2015-12-29 & 2015.1.01034.S & uid://A001/X2d8/X8b & false\\
  015017.70+002902.4 (1) & 01:50:17.71 & +00:29:02.5 & 3.010 & 01:50:17.81 & +00:29:03.8 & 2.06 & 7 & 7m & 3.31 & 0.706 & 6.9 & 11.4 & 20.8 & 12.8 & 9.3 & 3.7 & 23 & 4.5 & 2.8 & -89 & 0.61 & 2017-07-06 & 2016.2.00060.S & uid://A001/X124a/X171 & false\\
  021727.65-051502.8 (1) & 02:17:27.66 & -05:15:02.8 & 2.295 & 02:17:27.66 & -05:15:02.9 & 0.02 & 7 & 12m & 0.51 & 0.085 & 0.7 & 7.4 & 1.2 & 8.2 & 0.7 & 0.6 & 78 & 0.6 & 0.4 & -75 & 0.10 & 2018-08-24 & 2017.1.01492.S & uid://A001/X1284/X1752 & false\\
  021757.30-050808.6 (1) & 02:17:57.30 & -05:08:08.6 & 1.407 & 02:17:57.30 & -05:08:08.6 & 0.03 & 7 & 12m & 0.18 & 0.210 & 1.4 & 13.9 & 1.6 & 14.7 & 0.8 & 0.8 & 88 & 0.9 & 0.7 & 77 & 0.10 & 2018-06-06  & 2017.1.01027.S & uid://A001/X12a3/X36 & false\\
  074110.70+311200.2 (1) & 07:41:10.70 & +31:12:00.2 & 0.631 & 07:41:10.71 & +31:12:00.1 & 0.08 & 5 & 12m & 0.71 & 0.016 & 230.0 & 340.0 & 248.0 & 355.0 & 0.9 & 0.6 & 155 & 0.9 & 0.6 & -25 & 0.68 & 2018-09-05 & 2017.1.01559.S & uid://A001/X1296/Xaa9 & false\\
  095923.55+022227.2 (1) & 09:59:23.55 & +02:22:27.3 & 2.735 & 09:59:23.55 & +02:22:27.2 & 0.09 & 6 & 12m & 0.71 & 0.029 & 0.3 & 9.8 & 0.3 & 9.5 & 1.3 & 0.8 & 43 & 1.0 & 0.8 & 61 & 0.03 & 2018-12-21 & 2018.1.00681.S & uid://A001/X133d/X2262 & false\\
  100038.01+020822.4 (1) & 10:00:38.01 & +02:08:22.4 & 1.828 & 10:00:38.01 & +02:08:22.4 & 0.02 & 9 & 12m & 0.77 & 0.290 & 21.4 & 46.5 & 46.3 & 48.4 & 0.3 & 0.3 & 27 & 0.2 & 0.2 & 53 & 0.46 & 2016-11-16 & 2015.1.01362.S & uid://A001/X2d6/X236 & false\\
  101549.00+002020.0 (1) & 10:15:49.01 & +00:20:20.0 & 4.400 & 10:15:49.01 & +00:20:20.0 & 0.02 & 7 & 12m & 0.17 & 0.035 & 0.4 & 9.9 & 0.5 & 10.6 & 0.2 & 0.2 & 82 & 0.2 & 0.2 & 89 & 0.04 & 2016-11-03 & 2016.1.00718.S & uid://A001/X87d/X613 & false\\
  110045.23+112239.1 & 11:00:45.24 & +11:22:39.1 & 4.740 & 11:00:45.23 & +11:22:39.1 & 0.16 & 7 & 12m & 0.77 & 0.026 & 0.4 & 22.3 & 0.5 & 22.7 & 1.0 & 0.8 & 130 & 0.9 & 0.6 & -63 & 0.02 & 2017-03-22 & 2016.1.00569.S & uid://A001/X879/X219 & false\\
   &  &  &  & 11:00:44.91 & +11:22:40.1 (A) & 4.87 & 7 & 12m & 0.77 & 0.026 & 0.1 & 3.6 & 0.0 & 4.0 & 0.7 & 0.4 & 102 & 0.9 & 0.6 & -63 & 0.02 & 2017-03-22 & 2016.1.00569.S & uid://A001/X879/X219 & false\\
  111200.93+065530.1 (1) & 11:12:00.94 & +06:55:30.1 & 2.541 & 11:12:00.94 & +06:55:30.1 & 0.05 & 6 & 12m & 0.69 & 0.014 & 0.4 & 19.2 & 0.4 & 20.7 & 0.7 & 0.5 & 102 & 0.7 & 0.5 & -74 & 0.02 & 2019-04-11 & 2018.1.00583.S & uid://A001/X1374/X7a & false\\
  120110.31+211758.4 & 12:01:10.31 & +21:17:58.5 & 4.579 & 12:01:10.32 & +21:17:58.6 & 0.09 & 8 & 12m & 0.77 & 0.067 & 8.0 & 24.1 & 9.2 & 25.6 & 0.9 & 0.6 & 131 & 0.8 & 0.6 & -48 & 0.33 & 2015-12-29 & 2015.1.01564.S & uid://A001/X2f7/Xfe & false\\
   &  &  &  & 12:01:10.25 & +21:17:56.6 (A) & 2.10 & 8 & 12m & 0.20 & 0.040 & 0.1 & 4.5 & 0.2 & 4.9 & 0.5 & 0.2 & 149 & 0.3 & 0.2 & 2 & 0.03 & 2018-11-25 & 2017.1.01052.S & uid://A001/X1273/X766 & false\\
   &  &  &  & 12:01:10.29 & +21:17:55.9 (B) & 2.56 & 8 & 12m & 0.20 & 0.040 & 0.1 & 4.7 & 0.4 & 4.8 & 0.4 & 0.4 & 158 & 0.3 & 0.2 & 2 & 0.03 & 2018-11-25 & 2017.1.01052.S & uid://A001/X1273/X766 & false\\
   &  &  &  & 12:01:10.07 & +21:17:59.0 (C) & 3.40 & 8 & 12m & 0.20 & 0.195  & 0.1 & 3.8 & 0.2 & 4.1 & 0.4 & 0.3 & 107 & 0.3 & 0.2 & 2 & 0.03 & 2018-11-25 & 2017.1.01052.S & uid://A001/X1273/X766 & false\\
   &  &  &  & 12:01:10.07 & +21:17:59.0 (C) & 3.39 & 8 & 12m & 0.20 & 0.040 & 0.1 & 3.6 & 0.3 & 3.8 & 0.4 & 0.3 & 91 & 0.3 & 0.2 & 2 & 0.03 & 2018-11-25 & 2017.1.01052.S & uid://A001/X1273/X766 & false\\
    \hline
\end{tabular}}
\end{table}
\end{landscape}

\begin{landscape}

\begin{table}
\caption{Extractions of multiple images of the six lensed quasars. The catalogue contains the same columns as in Table \ref{tab:main}. 
}
\adjustbox{width=1.33\textwidth}{
\centering
\label{tab:lensed}
\begin{tabular}{lcccccccccclllllclccrrcclc}
    \hline
    \multicolumn{4}{c}{SDSS} &
    \multicolumn{2}{c}{ALMA} &
    Sep. &
    Band &
    Array &
    Ang. res. &
    Sens. &
    S$_{\rm peak}$ &
    S$_{\rm peak}$  &
    S$_{\rm int}$ &
    S$_{\rm int}$  &
    $\alpha_{max}$ &
    $\alpha_{min}$ &
    PA &
    $\alpha_{max}^{b}$ &
    $\alpha_{min}^{b}$ &
    PA$^{b}$ &
    RMS &
    Obs. date &
    Project code &
    MOUS ID &
    ALMACAL \\
    Name &
    RA &
    Dec &
    $z$ &
    RA &
    Dec &
    ($\arcsec$) &
     & 
     & 
    ($\arcsec$) &
    (mJy) &
    (mJy) &
    SNR &
    (mJy) &
    SNR &
    ($\arcsec$) &
    ($\arcsec$) &
    ($^{\circ}$) &
    ($\arcsec$) &
    ($\arcsec$) &
    ($^{\circ}$) &
    (mJy) &
     & 
     & 
     & 
     \\
    \hline
  081331.28+254503.0 & 08:13:31.29 & +25:45:03.1 & 1.510 & 08:13:31.28 & +25:45:03.1 & 0.07 & 4 & 12m & 0.83 & 0.030 & 0.2 & 6.2 & 0.3 & 7.6 & 1.6 & 1.0 & 162 & 1.2 & 0.8 & 19 & 0.03 & 2018-03-25 & 2017.1.01368.S & uid://A001/X1284/X2331 & false\\
   &   &   &   & 08:13:31.28 & +25:45:03.1 & 0.17 & 6 & 12m & 0.31 & 0.041 & 0.4 & 10.8 & 1.2 & 11.1 & 0.8 & 0.5 & 169 & 0.5 & 0.3 & 22 & 0.03 & 2018-09-14 & 2017.1.01081.S & uid://A001/X1273/X77e & false\\
   &   &   &   & 08:13:31.33 & +25:45:02.9 & 0.60 & 6 & 12m & 0.31 & 0.041 & 0.2 & 4.7 & 0.4 & 5.7 & 0.6 & 0.4 & 24 & 0.5 & 0.3 & 22 & 0.03 & 2018-09-14 & 2017.1.01081.S & uid://A001/X1273/X77e & false\\
   &   &   &   & 08:13:31.31 & +25:45:03.6 & 0.59 & 6 & 12m & 0.31 & 0.041 & 0.2 & 4.9 & 0.2 & 4.8 & 0.6 & 0.3 & 54 & 0.5 & 0.3 & 22 & 0.03 & 2018-09-14 & 2017.1.01081.S & uid://A001/X1273/X77e & false\\
  091127.61+055054.1 & 09:11:27.61 & +05:50:54.1 & 2.798 & 09:11:27.62 & +05:50:54.2 & 0.20 & 3 & 12m & 0.25 & 0.010 & 0.1 & 5.0 & 0.2 & 5.2 & 0.6 & 0.4 & 10 & 0.3 & 0.2 & -51 & 0.01 & 2019-08-19 & 2018.1.01008.S & uid://A001/X133d/X2882 & false\\
    &   &   &   & 09:11:27.41 & +05:50:54.7 & 3.01 & 4 & 12m & 0.26 & 0.012 & 0.1 & 5.8 & 0.2 & 5.8 & 0.4 & 0.3 & 142 & 0.3 & 0.2 & -90 & 0.02 & 2018-01-11 & 2017.1.01081.S & uid://A001/X1273/X792 & false\\
    &   &   &   & 09:11:27.62 & +05:50:54.3 & 0.26 & 6 & 12m & 1.03 & 0.046 & 7.2 & 109.0 & 10.4 & 116.0 & 1.3 & 1.0 & 22 & 1.1 & 0.9 & 54 & 0.07 & 2019-11-21 & 2019.1.01802.S & uid://A001/X1471/X25e & false\\
    &   &   &   & 09:11:27.63 & +05:50:54.2 & 0.21 & 4 & 12m & 0.26 & 0.012 & 0.3 & 17.8 & 1.0 & 17.4 & 0.6 & 0.4 & 19 & 0.3 & 0.2 & -90 & 0.02 & 2018-01-11 & 2017.1.01081.S & uid://A001/X1273/X792 & false\\
    &   &   &   & 09:11:27.63 & +05:50:54.3 & 0.27 & 7 & 12m & 0.08 & 0.037 & 0.6 & 12.6 & 5.5 & 13.0 & 0.3 & 0.2 & 38 & 0.1 & 0.1 & -70 & 0.05 & 2019-09-01 & 2018.1.01008.S & uid://A001/X133d/X288a & false\\
    &   &   &   & 09:11:27.42 & +05:50:54.6 & 2.99 & 7 & 12m & 0.08 & 0.037 & 0.3 & 6.6 & 2.6 & 6.3 & 0.3 & 0.2 & 76 & 0.1 & 0.1 & -70 & 0.05 & 2019-09-01 & 2018.1.01008.S & uid://A001/X133d/X288a & false\\
  092455.79+021924.9 & 09:24:55.83 & +02:19:25.0 & 1.522 & 09:24:55.81 & +02:19:25.3 & 0.45 & 6 & 12m & 0.69 & 0.047 & 0.5 & 4.8 & 0.5 & 4.3 & 0.9 & 0.6 & 125 & 0.7 & 0.6 & 74 & 0.10 & 2018-12-19 & 2018.1.01447.S & uid://A001/X133d/X3d4d & false\\
    &   &   &   & 09:24:55.81 & +02:19:25.4 & 0.44 & 7 & 12m & 0.17 & 0.027 & 0.5 & 15.8 & 1.0 & 15.3 & 0.3 & 0.2 & 103 & 0.2 & 0.2 & 8 & 0.03 & 2018-10-20 & 2018.1.01591.S & uid://A001/X133d/X3f38 & false\\
    &   &   &   & 09:24:55.85 & +02:19:25.0 & 0.30 & 7 & 12m & 0.17 & 0.027 & 0.4 & 11.3 & 0.8 & 10.1 & 0.3 & 0.2 & 154 & 0.2 & 0.2 & 8 & 0.03 & 2018-10-20 & 2018.1.01591.S & uid://A001/X133d/X3f38 & false\\
    &   &   &   & 09:24:55.75 & +02:19:24.7 & 1.26 & 7 & 12m & 0.17 & 0.027 & 0.3 & 9.8 & 0.9 & 10.2 & 0.4 & 0.2 & 26 & 0.2 & 0.2 & 8 & 0.03 & 2018-10-20 & 2018.1.01591.S & uid://A001/X133d/X3f38 & false\\
    &   &   &   & 09:24:55.82 & +02:19:23.6 & 1.42 & 7 & 12m & 0.17 & 0.027 & 0.3 & 7.5 & 1.0 & 8.1 & 0.4 & 0.3 & 66 & 0.2 & 0.2 & 8 & 0.03 & 2018-10-20 & 2018.1.01591.S & uid://A001/X133d/X3f38 & false\\
  111816.94+074558.2 & 11:18:16.95 & +07:45:58.2 & 1.735 & 11:18:16.86 & +07:46:00.0 & 2.29 & 7 & 12m & 0.20 & 0.034 & 0.2 & 6.6 & 0.3 & 7.5 & 0.3 & 0.2 & 21 & 0.2 & 0.2 & 64 & 0.04 & 2018-10-24 & 2018.1.01591.S & uid://A001/X133d/X3f3c & false\\
    &   &   &   & 11:18:16.95 & +07:45:58.0 & 0.15 & 7 & 12m & 0.20 & 0.034 & 0.7 & 21.0 & 1.1 & 20.6 & 0.3 & 0.2 & 37 & 0.2 & 0.2 & 64 & 0.04 & 2018-10-24 & 2018.1.01591.S & uid://A001/X133d/X3f3c & false\\
    &   &   &   & 11:18:16.96 & +07:45:58.5 & 0.29 & 7 & 12m & 0.20 & 0.034 & 0.7 & 19.7 & 1.3 & 19.0 & 0.4 & 0.2 & 22 & 0.2 & 0.2 & 64 & 0.04 & 2018-10-24 & 2018.1.01591.S & uid://A001/X133d/X3f3c & false\\
    &   &   &   & 11:18:16.84 & +07:45:58.1 & 1.67 & 7 & 12m & 0.20 & 0.034 & 0.2 & 5.8 & 0.3 & 6.6 & 0.3 & 0.2 & 97 & 0.2 & 0.2 & 64 & 0.04 & 2018-10-24 & 2018.1.01591.S & uid://A001/X133d/X3f3c & false\\
  133018.64+181032.1 & 13:30:18.65 & +18:10:32.2 & 1.393 & 13:30:18.65 & +18:10:32.1 & 0.10 & 7 & 12m & 0.34 & 0.025 & 0.4 & 13.9 & 1.2 & 14.7 & 0.7 & 0.4 & 88 & 0.4 & 0.3 & -29 & 0.03 & 2020-02-29 & 2019.1.00948.S & uid://A001/X1465/X1f70 & false\\
    &   &   &   & 13:30:18.58 & +18:10:33.2 & 1.45 & 7 & 12m & 0.34 & 0.025 & 0.2 & 7.1 & 0.8 & 6.7 & 0.8 & 0.6 & 165 & 0.4 & 0.3 & -29 & 0.03 & 2020-02-29 & 2019.1.00948.S & uid://A001/X1465/X1f70 & false\\
    &   &   &   & 13:30:18.69 & +18:10:33.7 & 1.59 & 7 & 12m & 0.34 & 0.025 & 0.2 & 6.4 & 0.9 & 6.0 & 1.3 & 0.4 & 150 & 0.4 & 0.3 & -29 & 0.03 & 2020-02-29 & 2019.1.00948.S & uid://A001/X1465/X1f70 & false\\
  141546.24+112943.4 & 14:15:46.24 & +11:29:43.4 & 2.560 & 14:15:46.25 & +11:29:43.6 & 0.20 & 3 & 12m & 2.15 & 0.014 & 0.4 & 19.1 & 0.5 & 20.8 & 2.6 & 1.9 & 68 & 2.7 & 1.6 & 61 & 0.02 & 2016-04-08 & 2015.1.01309.S & uid://A001/X2fb/X140 & false\\
    &   &   &   & 14:15:46.25 & +11:29:44.2 & 0.77 & 7 & 12m & 0.17 & 0.053 & 1.1 & 10.3 & 2.7 & 10.0 & 0.3 & 0.3 & 117 & 0.2 & 0.2 & 9 & 0.11 & 2015-06-30 & 2012.1.00175.S & uid://A001/X13b/X27 & false\\
    &   &   &   & 14:15:46.23 & +11:29:43.1 & 0.39 & 7 & 12m & 0.21 & 0.059 & 3.7 & 83.0 & 8.5 & 84.5 & 0.5 & 0.2 & 118 & 0.3 & 0.2 & -37 & 0.05 & 2015-06-25 & 2012.1.00175.S & uid://A001/X13b/X37 & false\\
    &   &   &   & 14:15:46.23 & +11:29:43.1 & 0.35 & 3 & 12m & 0.36 & 0.011 & 0.1 & 17.5 & 0.2 & 16.9 & 0.7 & 0.5 & 96 & 0.4 & 0.4 & -85 & 0.01 & 2017-12-21 & 2017.1.01232.S & uid://A001/X1288/X6be & false\\
    &   &   &   & 14:15:46.27 & +11:29:43.3 & 0.48 & 3 & 12m & 0.36 & 0.011 & 0.1 & 17.7 & 0.2 & 17.7 & 0.7 & 0.5 & 43 & 0.4 & 0.4 & -85 & 0.01 & 2017-12-21 & 2017.1.01232.S & uid://A001/X1288/X6be & false\\
    &   &   &   & 14:15:46.20 & +11:29:43.8 & 0.75 & 3 & 12m & 0.36 & 0.011 & 0.1 & 14.4 & 0.2 & 13.5 & 0.7 & 0.6 & 164 & 0.4 & 0.4 & -85 & 0.01 & 2017-12-21 & 2017.1.01232.S & uid://A001/X1288/X6be & false\\
    &   &   &   & 14:15:46.25 & +11:29:43.2 & 0.27 & 3 & 12m & 0.44 & 0.013 & 0.1 & 19.6 & 0.7 & 21.1 & 2.2 & 0.8 & 88 & 0.6 & 0.6 & -66 & 0.01 & 2018-01-12 & 2017.1.01232.S & uid://A001/X1288/X6c2 & false\\
    &   &   &   & 14:15:46.27 & +11:29:43.3 & 0.51 & 4 & 12m & 0.17 & 0.032 & 0.2 & 5.9 & 0.9 & 5.0 & 0.5 & 0.3 & 146 & 0.2 & 0.1 & -56 & 0.04 & 2017-12-06 & 2017.1.01081.S & uid://A001/X1296/Xc7b & false\\
    &   &   &   & 14:15:46.23 & +11:29:43.1 & 0.39 & 4 & 12m & 0.17 & 0.032 & 0.2 & 5.5 & 0.7 & 6.0 & 0.4 & 0.2 & 105 & 0.2 & 0.1 & -56 & 0.04 & 2017-12-06 & 2017.1.01081.S & uid://A001/X1296/Xc7b & false\\
    &   &   &   & 14:15:46.26 & +11:29:43.1 & 0.46 & 4 & 12m & 0.17 & 0.032 & 0.2 & 5.7 & 0.8 & 5.1 & 0.4 & 0.3 & 131 & 0.2 & 0.1 & -56 & 0.04 & 2017-12-06 & 2017.1.01081.S & uid://A001/X1296/Xc7b & false\\
    &   &   &   & 14:15:46.19 & +11:29:43.7 & 0.75 & 4 & 12m & 0.17 & 0.032 & 0.2 & 4.5 & 0.4 & 4.5 & 0.4 & 0.2 & 115 & 0.2 & 0.1 & -56 & 0.04 & 2017-12-06 & 2017.1.01081.S & uid://A001/X1296/Xc7b & false\\
    &   &   &   & 14:15:46.25 & +11:29:44.1 & 0.70 & 4 & 12m & 0.17 & 0.032 & 0.2 & 4.0 & 0.4 & 4.4 & 0.3 & 0.3 & 148 & 0.2 & 0.1 & -56 & 0.04 & 2017-12-06 & 2017.1.01081.S & uid://A001/X1296/Xc7b & false\\
    &   &   &   & 14:15:46.26 & +11:29:43.3 & 0.32 & 4 & 12m & 0.59 & 0.017 & 0.3 & 14.2 & 0.8 & 14.5 & 1.0 & 0.8 & 49 & 0.6 & 0.5 & -40 & 0.02 & 2019-10-06 & 2019.1.00883.S & uid://A001/X1465/X20a9 & false\\
    &   &   &   & 14:15:46.20 & +11:29:43.8 & 0.73 & 4 & 12m & 0.59 & 0.017 & 0.2 & 11.4 & 0.7 & 12.1 & 1.1 & 0.7 & 153 & 0.6 & 0.5 & -40 & 0.02 & 2019-10-06 & 2019.1.00883.S & uid://A001/X1465/X20a9 & false\\
    &   &   &   & 14:15:46.26 & +11:29:43.3 & 0.27 & 4 & 12m & 0.71 & 0.019 & 0.5 & 23.7 & 1.2 & 25.2 & 1.3 & 1.0 & 56 & 0.8 & 0.7 & 18 & 0.02 & 2019-10-08 & 2019.1.00883.S & uid://A001/X1465/X20ad & false\\
    &   &   &   & 14:15:46.23 & +11:29:43.1 & 0.34 & 7 & 12m & 0.17 & 0.053 & 1.6 & 15.2 & 6.8 & 15.6 & 0.6 & 0.3 & 103 & 0.2 & 0.2 & 9 & 0.11 & 2015-06-30 & 2012.1.00175.S & uid://A001/X13b/X27 & false\\
    &   &   &   & 14:15:46.28 & +11:29:43.3 & 0.56 & 7 & 12m & 0.17 & 0.053 & 1.6 & 14.7 & 8.1 & 14.9 & 0.5 & 0.3 & 45 & 0.2 & 0.2 & 9 & 0.11 & 2015-06-30 & 2012.1.00175.S & uid://A001/X13b/X27 & false\\
    &   &   &   & 14:15:46.20 & +11:29:43.9 & 0.80 & 7 & 12m & 0.17 & 0.053 & 1.2 & 11.6 & 5.7 & 11.8 & 0.5 & 0.3 & 1 & 0.2 & 0.2 & 9 & 0.11 & 2015-06-30 & 2012.1.00175.S & uid://A001/X13b/X27 & false\\
    &   &   &   & 14:15:46.22 & +11:29:43.1 & 0.41 & 7 & 12m & 0.11 & 0.167 & 1.1 & 8.6 & 6.3 & 7.1 & 0.3 & 0.2 & 128 & 0.1 & 0.1 & 52 & 0.13 & 2015-09-27 & 2012.1.00175.S & uid://A001/X13b/X2b & false\\
    &   &   &   & 14:15:46.19 & +11:29:44.0 & 0.92 & 7 & 12m & 0.11 & 0.167 & 0.6 & 4.4 & 8.6 & 5.6 & 0.6 & 0.2 & 177 & 0.1 & 0.1 & 52 & 0.13 & 2015-09-27 & 2012.1.00175.S & uid://A001/X13b/X2b & false\\
    &   &   &   & 14:15:46.28 & +11:29:43.4 & 0.53 & 7 & 12m & 0.11 & 0.167 & 0.7 & 5.8 & 9.5 & 7.1 & 0.6 & 0.2 & 16 & 0.1 & 0.1 & 52 & 0.13 & 2015-09-27 & 2012.1.00175.S & uid://A001/X13b/X2b & false\\
    &   &   &   & 14:15:46.24 & +11:29:43.4 & 0.09 & 7 & 12m & 0.81 & 0.087 & 13.9 & 107.0 & 38.5 & 110.0 & 1.5 & 1.1 & 82 & 0.8 & 0.8 & 35 & 0.13 & 2018-05-22 & 2017.1.00963.S & uid://A001/X12d1/X315 & false\\
    \hline
\end{tabular}}
\end{table}
\end{landscape}

\begin{table*}
\caption{The first 20 lines in the catalogue of SDSS quasars that were not detected by ALMA. The columns are SDSS Name, RA and Dec; redshift ($z$); ALMA band; angular resolution (Ang. res.) in $\arcsec$; continuum sensitivity (Sens.) in mJy, project code and MOUS ID.}
\adjustbox{width=0.99\textwidth}{
\centering
\label{tab:NonDetected}
\begin{tabular}{lcccclcccl}
    \hline
    \multicolumn{4}{c}{SDSS} &
    Band &
    Array &
    Ang. res. &
    Sens. &
    Project code &
    MOUS ID \\
    Name &
    RA &
    Dec &
    $z$ &
     &
     &
    ($\arcsec$) &
    (mJy) &
     & \\
    \hline
  00000-00000 & 02:30:24.47 & -04:09:13.4 & 2.241 & 3 & 12m & 1.02 & 0.020 & 2016.1.00798.S & uid://A001/X87d/X741\\
  000746.19+122223.9 & 00:07:46.20 & +12:22:23.9 & 2.429 & 6 & 12m & 0.49 & 0.014 & 2017.1.00478.S & uid://A001/X12d1/X1ba\\
  000826.77-050636.4 & 00:08:26.77 & -05:06:36.5 & 2.172 & 7 & 7m & 3.44 & 0.963 & 2016.2.00060.S & uid://A001/X124a/X16e\\
  001602.40-001225.1 & 00:16:02.40 & -00:12:25.1 & 2.085 & 4 & 12m & 1.68 & 0.011 & 2017.1.01558.S & uid://A001/X1289/X238\\
  002235.96+001850.0 & 00:22:35.96 & +00:18:50.0 & 0.522 & 7 & 12m & 0.10 & 0.065 & 2016.1.01523.S & uid://A001/X87c/X293\\
  002241.14+143112.6 & 00:22:41.15 & +14:31:12.6 & 2.729 & 3 & 12m & 0.08 & 0.014 & 2016.1.01231.S & uid://A001/X1262/X36\\
   &  &  &  & 6 & 12m & 1.15 & 0.038 & 2013.1.00262.S & uid://A001/X145/Xa2\\
   &  &  &  & 3 & 12m & 1.68 & 0.011 & 2019.1.00790.S & uid://A001/X1467/X2af\\
   &  &  &  & 4 & 12m & 1.17 & 0.018 & 2015.1.01157.S & uid://A001/X2f7/Xd5\\
   &  &  &  & 6 & 12m & 0.80 & 0.018 & 2015.1.01157.S & uid://A001/X2f7/Xdd\\
  002245.15+001822.5 & 00:22:45.15 & +00:18:22.5 & 1.754 & 7 & 12m & 0.10 & 0.065 & 2016.1.01523.S & uid://A001/X87c/X293\\
  003703.73+011707.6 & 00:37:03.73 & +01:17:07.6 & 2.198 & 3 & 12m & 1.76 & 0.028 & 2019.1.01257.S & uid://A001/X1467/X68\\
   &  &  &  & 3 & 12m & 1.70 & 0.032 & 2019.1.01257.S & uid://A001/X1467/X87\\
   &  &  &  & 3 & 12m & 1.63 & 0.031 & 2019.1.01257.S & uid://A001/X1467/Xa6\\
   &  &  &  & 3 & 12m & 1.50 & 0.036 & 2019.1.01257.S & uid://A001/X1467/Xc5\\
   &  &  &  & 3 & 12m & 1.47 & 0.037 & 2019.1.01257.S & uid://A001/X1467/Xe4\\
  003705.89+011611.4 & 00:37:05.90 & +01:16:11.4 & 0.651 & 3 & 12m & 1.76 & 0.028 & 2019.1.01257.S & uid://A001/X1467/X68\\
   &  &  &  & 3 & 12m & 1.70 & 0.032 & 2019.1.01257.S & uid://A001/X1467/X87\\
   &  &  &  & 3 & 12m & 1.63 & 0.031 & 2019.1.01257.S & uid://A001/X1467/Xa6\\
   &  &  &  & 3 & 12m & 1.50 & 0.036 & 2019.1.01257.S & uid://A001/X1467/Xc5\\
\hline
\end{tabular}}
\end{table*}

\begin{table*}
\caption{The four objects whose images did not have a primary beam response curve and for which the automatic flux extraction did not return any flux measurement. The columns are as in Table \ref{tab:NonDetected}.}
\adjustbox{width=0.99\textwidth}{
\centering
\label{tab:pbc}
\begin{tabular}{lccccccccl}
    \hline
    \multicolumn{4}{c}{SDSS} &
    Band &
    Array &
    Ang. res. &
    Sens. &
    Project code &
    MOUS ID \\
    Name &
    RA &
    Dec &
    $z$ &
     &
     &
    ($\arcsec$) &
    (mJy) &
     & \\
    \hline
  094932.26+033531.7 & 09:49:32.27 & +03:35:31.8 & 4.107 & 9 & 12m & 0.14 & 1.088 & 2012.1.00604.S &  uid://A002/X5ce05d/X10d\\
  100711.80+053208.8 & 10:07:11.81 & +05:32:09.0 & 2.148 & 9 & 12m & 0.14 & 0.993 & 2012.1.00604.S   & uid://A002/X5ce05d/X10d\\
  131341.19+144140.5 & 13:13:41.20 & +14:41:40.6 & 1.897 & 9 & 12m & 0.18 & 0.269 & 2013.1.00526.S & uid://A001/X11e/Xf\\
  135559.03-002413.6 & 13:55:59.04 & -00:24:13.7 & 2.332 & 6 & 12m & 0.45 & 0.064 & 2018.1.01817.S & uid://A001/X137b/X59\\
\hline
\end{tabular}}
\end{table*}


\bsp	
\label{lastpage}
\end{document}